\documentclass[a4]{article}
\usepackage{amsmath,amsfonts,setspace,cancel,url,hyperref,verbatim,soul,cite,xcolor}
\usepackage{fullpage}
\usepackage{breqn}
\usepackage[utf8]{inputenc}

\usepackage{tikz}
\newcommand{\TM}{}

\newcommand{\be}{\begin{equation}}
\newcommand{\ee}{\end{equation}}
\newcommand{\beqa}{\begin{eqnarray} }
\newcommand{\eeqa}{\end{eqnarray} }
\newcommand{\barray}{\begin{array}}
\newcommand{\earray}{\end{array}}

\newcommand{\bx}{\bar{x}}
\newcommand{\by}{\bar{y}}

\newcommand{\gM}{\mathcal{M}}

\newcommand\cH{{\cal H}}

\newcommand{\Gthree}{\mathrm{SL}(3) \times \mathrm{SL}(2)}

\numberwithin{equation}{section}

\newcommand{\bbeta}{\boldsymbol{\beta}}
\newcommand{\bbetabar}{\boldsymbol{\bar\beta}}

\makeatletter
\let\oldr@@t\r@@t
\def\r@@t#1#2{%
\setbox0=\hbox{$\oldr@@t#1{#2\,}$}\dimen0=\ht0
\advance\dimen0-0.2\ht0
\setbox2=\hbox{\vrule height\ht0 depth -\dimen0}%
{\box0\lower0.4pt\box2}}
\makeatother
\usepackage{letltxmacro}
\LetLtxMacro{\oldsqrt}{\sqrt}
\newcommand*{\sqrtbig}[2][\ ]{\oldsqrt[#1]{#2}}

\begin{document}

\title{\vspace{-2.5em}\bf Non-relativistic duality and $T \bar T$ deformations}

\author{\sc Chris D. A. Blair}

\date{
Theoretische Natuurkunde, Vrije Universiteit Brussel, and the International Solvay Institutes,  Pleinlaan 2, B-1050 Brussels, Belgium 
\\ {\tt cblair@vub.ac.be}
}

\maketitle

\begin{abstract}
We make some observations connecting non-relativistic limits of string theory with $T\bar T$ deformations and TsT transformations. 
\end{abstract}

\tableofcontents

\section{Deformations, actions and spectra} 

\subsection{Moving in the space of physical theories}

To orient ourselves in the space of physical theories, it can be useful to think of $c,G$ and $\hbar$ as parameters that tell us whether we are in a regime which is relativistic/non-relativistic, gravitational/non-gravitational or quantum/classical. Most of the combinations of these descriptors apply to theories which are familiar to all physicists. 

Another way to move in theory space is to start with a known theory and deform it by adding to the action a coupling to some operator in the original theory.
A surprising example of this is the $T\bar T$ deformation \cite{Zamolodchikov:2004ce,Smirnov:2016lqw, Cavaglia:2016oda}, see the lectures \cite{Jiang:2019hxb} for a nice introduction, which deforms a two-dimensional QFT using a coupling to the determinant of the energy-momentum tensor (hence the name). Although this deformation is irrelevant, it turns out to be unreasonably well-behaved as we go to the UV (corresponding to sending the coupling, or deformation parameter, $\lambda \rightarrow \infty$). For instance, if we know the original spectrum we can obtain the deformed one, while, remarkably, integrability of the original theory is preserved.

This deformation has turned out to have a direct link to string theory. The $T\bar T$ deformation of a theory of $D$ free bosons corresponds to the Nambu-Goto action in $D+2$ dimensions, with the two extra directions fixed to static gauge \cite{Cavaglia:2016oda, Bonelli:2018kik}.
As well as the usual Nambu-Goto square root term, we also need a non-zero $B$-field with a component in the longitudinal direction proportional to $1/\lambda$. 
Both this $B$-field and the Nambu-Goto square root are naively divergent in the limit $\lambda \rightarrow 0$, but these divergences cancel such that we recover the original undeformed theory. 

String theory (or more broadly its M-theoretic completion) is meant to occupy the position in $c,G, \hbar$ space corresponding to relativistic quantum gravity. 
Starting there, and thinking about moving in all possible directions of the $c,G,\hbar$ cube, we might wonder about the limit of string theory when $c^2 \rightarrow \infty$, which should correspond to non-relativistic quantum gravity. This limit may exhibit novel features of string theory, or more speculatively provide an alternative route to insights into quantum gravity more generally. Either way, understanding this corner of theory space has been a motivation for recent progress in \emph{non-relativistic string theory}.

The direct way to obtain a non-relativistic string theory is to take the string sigma model in a background spacetime and perform a scaling limit which treats the longitudinal time
and spatial directions of the string separately to the transverse ones \cite{Gomis:2000bd, Danielsson:2000gi,  Danielsson:2000mu,Andringa:2012uz} .
Effectively, the longitudinal components of the metric should scale like $c^2$,\footnote{For a point particle, only the time coordinate need be scaled, in which case this sort of limit is directly related to sending the speed of light to infinity. For branes, it is necessary to scale all worldvolume directions, in which case the parameter one is sending to infinity should not be directly thought of as the speed of light \cite{Gomis:2004pw}. For that reason we will denote the actual parameter that we send to infinity by $\omega^2$ below.} and to obtain a finite result, the $B$-field to which the string couples needs to have a longitudinal component also proportional to $c^2$. The naive divergence of both the metric and the $B$-field then cancel in the limit.

This scaling limit, with cancellation of divergences between metric and $B$-field contributions, should sound similar to what happens in the $T \bar T$ deformation as we return to the initial $\lambda = 0$ undeformed theory.
Working out how to make this connection explicit is the main goal of this paper.

The first part of our observations will focus on the scaling limit we have mentioned, working directly with Nambu-Goto action (in section \ref{nglimit}) and also with the spectrum (in section \ref{spectra}).

The second part concerns the geometrical viewpoint.
Interpreting the $T \bar T$ deformation in terms of a string worldsheet theory, we can ask whether there is a target space perspective.
One is provided by realising the effect of the deformation in terms of TsT transformations \cite{Lunin:2005jy} of a string theory geometry \cite{Baggio:2018gct, Araujo:2018rho, Frolov:2019nrr,Frolov:2019xzi,Sfondrini:2019smd}, working in the Hamiltonian formulation of the string (see also \cite{Jorjadze:2020ili}). The TsT transformations act on the two additional longitudinal transformations by which we extend the theory we wish to deform.
We will be able to extend this picture to deformed geometries which are singular in the limit $\lambda = 0$ (owing to the need to carry out one of the T-duality transformations of the TsT in a null direction \cite{Araujo:2018rho, Sfondrini:2019smd}), and interpret the apparently singular background describing the undeformed theory as a non-relativistic geometry.

The prototypical non-relativistic background for us is known as Newton-Cartan geometry.
In a Newton-Cartan geometry the non-degenerate spacetime metric is replaced by a pair of orthogonal degenerate metrics, one for the timelike direction and one for spatial hypersurfaces. The combination of names signifies that the geometry is non-relativistic (Newton) but geometric i.e. generally covariant (Cartan). One way to think of this is as general relativity with Galileo substituted for Lorentz.
As the string is an extended object, the coordinates corresponding both to the time and a single spatial direction of the target space are treated on a different footing to the other spatial coordinates, and in the worldsheet theory couple differently to the geometry.

The version of Newton-Cartan which arises naturally via the $c^2 \rightarrow \infty$ scaling limit is known as stringy Newton-Cartan geometry \cite{Andringa:2012uz}.
An alternative route to non-relativistic stringy geometries is providing by carrying out a T-duality transformation on a null isometry direction of a relativistic background (for fixed null momentum). This leads to what is called torsional Newton-Cartan geometry \cite{Harmark:2017rpg, Harmark:2018cdl}.
Similarly, starting with the stringy Newton-Cartan geometry constructed in \cite{Andringa:2012uz}, one can T-dualise on on a longitudinal isometry to arrive at a relativistic geometry with a null isometry \cite{Bergshoeff:2018yvt} (and in this case one can directly relate torsional Newton-Cartan to the string Newton-Cartan \cite{Harmark:2019upf}).

In order to treat these geometries on the same footing as conventional relativistic geometries, and to relate them via null T-duality transformations, we will use ideas that were developed in \cite{Lee:2013hma, Ko:2015rha} and then systematically in \cite{Morand:2017fnv} in the context of the ``doubled'' (or, loosely speaking, T-duality covariant) formulation of string theory.
For the purposes of this paper, the main idea can be found in the Hamiltonian formulation of the string, where the metric and $B$-field appear together in an $O(D,D)$-valued matrix which we call the generalised metric. We first discuss this in section \ref{limham}. Whereas under the conventional radial inversion duality transformations, the metric and $B$-field transform non-linearly, the generalised metric transforms merely via a permutation of its entries. Therefore working with this object evades the singularities that would otherwise arise in null dualities, and via \cite{Morand:2017fnv} can be reparametrised in non-relativistic backgrounds in a geometrically transparent manner. We discuss this in section \ref{nonrieparam}.

Using this language, we will be able to complete the circle of ideas, first relating non-relativistic scaling limits to $T \bar T$ deformations at the level of the Nambu-Goto action, and then understanding this geometrically in string theory as an $O(D,D)$ transformation that makes the non-relativistic target space into a relativistic one.

In the remainder of section \ref{two} we will obtain some more general statements of our results from this first section, and discuss some particular examples of deformations relating non-relativistic and relativistic geometries. 
In section \ref{mdef} we discuss how all this does, and does not, generalise to M-theory.
Finally, rather than conclude in the conventional style with a conclusion, in this paper we will pause at the end of each section to provide a summation of the main ideas that have appeared so far.

\subsection{Limits of the Nambu-Goto action}
\label{nglimit}

Take a background which is just flat space with a constant $B$-field:
\be
\begin{split}
 ds^2 & = \omega^2 ( - (dX^0)^2 + (dX^1)^2 ) + \delta_{ij} dX^i dX^j \,,\\
 B_{01} & = \omega^2 \,,
\end{split}
\label{initial}
\ee
where we have rescaled two of the coordinates by a (dimensionless) constant $\omega^2$, and here $i,j=2,\dots,D$ with $D=10$ or $26$ the critical dimension.

Our first goal is to take the $\omega^2 \rightarrow \infty$ limit in the Nambu-Goto action in static gauge.
This is the non-relativistic limit of \cite{Gomis:2000bd, Danielsson:2000gi}, although we are expressing it slightly differently. 
In appendix \ref{unitz} we are a little more precise about the identification (specifically with the conventions of \cite{Gomis:2000bd}). In particular, one can think of this limit as being a zero slope limit with a critical $B$-field, identifying $\omega^2 =\frac{\alpha^\prime_{\text{eff}}}{\alpha^\prime}$ and sending $\alpha^\prime \rightarrow 0$. The actions that appear below should then be multiplied by an effective tension $T_{\text{eff}} = \frac{1}{2 \pi \alpha^\prime_{\text{eff}}}$, which we however pre-emptively set to unity.

The worldsheet coordinates are $\sigma^\alpha=(\tau,\sigma)$ and derivatives with respect to these coordinates will be denoted by $\dot{X}^\mu \equiv \partial_\tau X^\mu$, $X^{\prime \mu} \equiv \partial_\sigma X^\mu$. We also use the two-dimensional alternating symbol $\epsilon^{\alpha \beta} = -\epsilon^{\beta \alpha}$ with $\epsilon^{01} = -1$. The Nambu-Goto action coupled to a general metric $g_{\mu\nu}$ and $B$-field $B_{\mu\nu}$ is:
\be
S_{\text{NG}} = - 
 \int d^2\sigma \sqrt{- \det g_{\alpha \beta}} 
 -
 \int d^2 \sigma \frac{1}{2} \epsilon^{\alpha \beta} B_{\alpha \beta} \,,
\label{NG}
\ee
where the pullbacks of the metric and $B$-field are $g_{\alpha \beta} \equiv \partial_\alpha X^\mu \partial_\beta X^\nu g_{\mu\nu}$ and $B_{\alpha \beta} \equiv \partial_\alpha X^\mu \partial_\beta X^\nu B_{\mu\nu}$.

We take $X^0 =\tau$ and $X^1 = \sigma$ to fix static gauge.
The worldsheet action \eqref{NG} in the background \eqref{initial} then becomes:
\be
S_{\text{NG}}\Big|_{\text{static gauge}} = 
\int d^2 \sigma \, \omega^2 \left(1 
- \sqrtbig{
1 - \frac{1}{\omega^2} ( \dot{X}^i \dot{X}^j - X^{\prime i} X^{\prime j}) \delta_{ij} - \frac{1}{\omega^4} \det ( \partial_\alpha X^i \partial_\beta X^j \delta_{ij} )
}\right) \,.
\label{ngsg}
\ee
Take $\omega^2 \rightarrow \infty$. We can expand 
\be
S_{\text{NG}}\Big|_{\text{static gauge},\, \omega^2 \rightarrow \infty} = 
S_0 + \frac{1}{\omega^2} S_1 + O \left(\frac{1}{\omega^4} \right) \,,
\ee
where the term that survives in the strict $\omega^2 \rightarrow \infty$ limit is just the action for $D-2$ free bosons:
\be
S_0 = 
 \int d^2\sigma \frac{1}{2} ( \dot{X}^i \dot{X}^j - X^{\prime i} X^{\prime j}) \delta_{ij} \,,
\label{S0}
\ee
and the first correction for finite $\omega$ involves the determinant of the energy-momentum tensor of $S_0$:
\be
S_1 =
\int d^2\sigma
\frac{1}{2} 
\left( \det (\partial_\alpha X^i \partial_\beta X^j \delta_{ij} )
+ \frac{1}{4} \left( \left( \dot{X}^i \dot{X}^j - X^{\prime i} X^{\prime j}\right) \delta_{ij}\right)^2 
\right)
= 
\frac{1}{2}
\int d^2\sigma \det (T_{\alpha \beta} )\,,
\ee
with
\be
T_{\alpha \beta} =\partial_\alpha X^i \partial_\beta X^j \delta_{ij} - \frac{1}{2} \eta_{\alpha \beta} \eta^{\gamma\delta} \partial_\gamma X^i \partial_\delta X^j \delta_{ij} \,,
\ee
where $\eta_{\alpha \beta} = \mathrm{diag}(-1,1)$ is the two-dimensional Minkowski metric.
Thus we can think of turning on a finite $\omega^2$ as deforming the theory with action $S_0$ by the determinant of the energy-momentum tensor.
It is convenient to define our deformation parameter as the inverse of $\omega^2$, namely
\be
\lambda = \frac{1}{\omega^2} \,,
\ee
such that the undeformed theory corresponds to $\lambda = 0$. 
The dependence on $\lambda$ of the classical action of the deformed theory follows from an equation of the form \cite{Zamolodchikov:2004ce,Smirnov:2016lqw, Cavaglia:2016oda,Jiang:2019hxb}:
\be
\frac{\partial S(\lambda)}{\partial \lambda}= \frac{1}{2}\int d^2\sigma \det (T_{\alpha \beta}(\lambda))\,.
\label{SflowTTbar}
\ee
We offer a nice demonstration of this in appendix \ref{flow}.
Not only can we deform the classical action -- and obtain closed form results for the resulting theory -- but we can study features such as the spectrum and S-matrix, and track how they change with respect to $\lambda$ (although we will not really consider quantum aspects in this paper). 
Given the identifications of appendix \ref{unitz}, we realise that we can really identify in our conventions $\lambda = \frac{\alpha^\prime}{\alpha^\prime_{\text{eff}}}$ (or reinstating the effective tension, we actually have the dimensionful quantity $\lambda /T_{\text{eff}} \sim \alpha^\prime$, see appendix \ref{unitz}). The limit $\lambda \rightarrow 0$ is then a field theory limit (or derivative expansion) of the string as pointed out regarding the non-relativistic limit in \cite{Gomis:2004pw} and in line with the observations made even earlier in \cite{Townsend:1999hi}, as well as with the reverse intuition that $T \bar T$ deformations lead to non-local theories.
 
The example above shows that the effect of such a deformation can be encoded in a string theory geometry, a crucial feature being the interplay between the metric and the $B$-field.
The geometry of string theory is of course a very useful tool for encoding interesting physics in a variety of ways.
Indeed, the link between $T \bar T$ deformations and string theory extends beyond the simple example above, which connected a free field theory to string theory in flat space with a divergent $B$-field in static gauge. 
For more complicated field theories, described by other geometries, a precise link has been elucidated in  \cite{Baggio:2018gct,Araujo:2018rho, Frolov:2019nrr,Frolov:2019xzi,Sfondrini:2019smd} by describing the $T \bar T$ deformation in terms of the well-known TsT transformations which involve T-duality, a geometric shift (either of the coordinates or of the $B$-field) and then a second T-duality.
We will describe this in more detail later on. 
(Note that from the point of view of this general approach, one can avoid introducing divergent $B$-fields -- and view the appearance of such in the initial example as an artefact of flat space -- while the deformation is most clearly expressed not in static gauge but in uniform light cone gauge.)

To connect with this more general picture for our example geometry \eqref{initial}, we will change perspective.
We want to be able to chase the limit $\omega^2 \rightarrow \infty$, or $\lambda \rightarrow 0$, directly at the level of the geometry.
In terms of the metric and $B$-field in \eqref{initial}, this is inherently problematic. 
For an alternative viewpoint on the geometry, we turn to the Hamiltonian formulation of the string.

\subsection{Limits of the Hamiltonian action}
\label{limham}

Before directly discussing the background \eqref{initial} in Hamiltonian language, we will first set-up some general notation which is key to understanding the main arguments of this paper.

\subsubsection*{The Hamiltonian form of the worldsheet action}

By the Hamiltonian form of the worldsheet action, we mean the formulation in which our independent fields are the coordinates $X^\mu$ and their conjugate momenta $P_\mu$.
The action in this form is:
\be
\begin{split}
S = \int d\tau d\sigma \left( \dot{X}^\mu P_\mu - u \mathcal{H}_u - e \mathcal{H}_e \right) \,.
\label{SHAM}
\end{split} 
\ee
In \eqref{SHAM} we also have two Lagrange multipliers, $e$ and $u$, imposing constraints given by:
\be
\begin{split}
\mathcal{H}_u & = X^{\prime \mu} P_\mu 
\,,\\
\mathcal{H}_e & = \frac{1}{2} \begin{pmatrix} X^{\prime \mu} & P_\mu \end{pmatrix} 
\begin{pmatrix} 
g_{\mu\nu} - B_{\mu\rho} g^{\rho\sigma} B_{\sigma \nu} & B_{\mu\rho} g^{\rho \nu} \\
- g^{\mu\rho} B_{\rho \nu} & g^{\mu\nu}
\end{pmatrix}
\begin{pmatrix} 
X^{\prime \nu} \\ P_\nu
\end{pmatrix}\,.
\end{split} 
\label{HamConstraints}
\ee
On integrating out the momenta, we recover the Polyakov action after identifying $e$ and $u$ with the independent components of the worldsheet metric. Further integrating out $e$ and $u$ leads to the Nambu-Goto action. The presence of the two constraints corresponds then to the equations of motion of the worldsheet metric, that is to the vanishing of the energy-momentum tensor of the string (i.e. the Virasoro constraints).
These are first-class constraints and generate worldsheet diffeomorphisms as their gauge transformations.

\subsubsection*{The generalised metric}

The structure of these constraints warrants further attention.
Firstly, note we can rewrite $\mathcal{H}_u$ as
\be
\mathcal{H}_u 
= \frac{1}{2} \begin{pmatrix} X^{\prime \mu} & P_\mu \end{pmatrix}
\begin{pmatrix} 0 & \delta_\mu{}^\nu \\ \delta^\mu{}_\nu & 0 \end{pmatrix} 
\begin{pmatrix} 
X^{\prime \nu} \\ P_\nu
\end{pmatrix}\,.
\ee
Then in both $\mathcal{H}_u$ and $\mathcal{H}_e$ we see the appearance of a $2D \times 2D$ matrix.
That in $\mathcal{H}_u$ we will call 
\be
\eta_{MN} = \begin{pmatrix} 0 & \delta_\mu{}^\nu \\ \delta^\mu{}_\nu & 0 \end{pmatrix} \,,
\ee
defining a split signature bilinear form preserved by the group $O(D,D)$ (which is not a priori a symmetry of the worldsheet action at all). We denote its inverse by $\eta^{MN}$. 
Here we have introduced an index $M$ which is $2D$-dimensional, and splits into upper and lower $D$-dimensional indices. For instance, we define $\mathcal{Z}^M \equiv ( X^{\prime \mu} , P_\mu )$, with $\mathcal{H}_u = \frac{1}{2} \eta_{MN} \mathcal{Z}^M \mathcal{Z}^N$.

We similarly treat $\mathcal{H}_e$ by writing it as $\mathcal{H}_e = \frac{1}{2} \cH_{MN} \mathcal{Z}^M \mathcal{Z}^N$, where by definition i) $\mathcal{H}_{MN} = \mathcal{H}_{NM}$, ii) $\mathcal{H}_{MN} \eta^{NK} \mathcal{H}_{KL} = \eta_{ML}$, i.e. $\mathcal{H}_{MN}$ is itself valued in the group $O(D,D)$.
The matrix $\mathcal{H}_{MN}$ will be referred to as the \emph{generalised metric}.

We will say that a (classical) string Hamiltonian is defined by the above action and constraints, with $\mathcal{H}_{MN}$ defined as above.
In a well-defined \emph{relativistic} (or [pseudo-]\emph{Riemannian}) spacetime background we parametrise this matrix $\mathcal{H}_{MN}$ as in \eqref{HamConstraints}:
\be
\text{\underline{Riemannian background:}}\quad
\mathcal{H}_{MN} = 
\begin{pmatrix} 
g_{\mu\nu} - B_{\mu\rho} g^{\rho\sigma} B_{\sigma \nu} & B_{\mu\rho} g^{\rho \nu} \\
- g^{\mu\rho} B_{\rho \nu} & g^{\mu\nu}
\end{pmatrix}\,.
\label{GMRie}
\ee
This is a parametrisation of the coset $O(D,D) / O(1,D-1) \times O(1,D-1)$, but not the only useful one.

\subsubsection*{Hamiltonian description of the geometry \eqref{initial}}
  
Our initial geometry \eqref{initial} produces a Hamiltonian with interesting properties as we vary the parameter $\lambda = \frac{1}{\omega^2}$.
Let's compactify our notation by writing the coordinates as $X^\mu = (X^a, X^i)$, where $a=0,1$ and $i=2,\dots ,D$ as before.
Define
\be
\eta_{ab} = \begin{pmatrix} -1 & 0 \\ 0 & 1 \end{pmatrix} \,,\quad
\epsilon^{ab} = \begin{pmatrix} 0 & - 1 \\ 1& 0 \end{pmatrix} \,,\quad
\epsilon_a{}^b \equiv\eta_{ac} \epsilon^{cb}  = \begin{pmatrix} 0 & 1 \\ 1 & 0 \end{pmatrix} \,.
\ee
The background \eqref{initial} is then encoded by the following generalised metric
\be
\mathcal{H}_{MN}  = 
\begin{pmatrix}
0  & 0 & \epsilon_a{}^b & 0\\
0 & \delta_{ij} & 0 & 0 \\
\epsilon_b{}^a &0 &  \lambda \eta^{ab} & 0 \\
0 & 0 & 0 &\delta^{ij} \\
\end{pmatrix} \,,
\label{GMGO}
\ee
and the Hamiltonian form of the action is thus
\be
\begin{split}
S = \int d^2\sigma \Big(&
\dot{X}^a P_a + \dot{X}^i P_i- u \left( X^{\prime a} P_a + X^{\prime i} P_i \right)
\\ & 
- \frac{e}{2} \left(
\delta^{ij}P_iP_j + \delta_{ij} X^{\prime i} X^{\prime j} 
 + 2 \epsilon_b{}^a P_a X^{\prime b} 
 + \lambda \eta^{ab}P_a P_b
\right)
\Big)\,.
\end{split}
\label{SHGO}
\ee
For $\lambda \rightarrow 0$, the metric and $B$-field of \eqref{initial} are singular.
The generalised metric \eqref{GMGO} and the action \eqref{SHGO} suffer no divergences.
Instead, the singularity manifests itself in the bottom right block of the matrix \eqref{GMGO} becoming degenerate -- this was the block corresponding to the inverse spacetime metric -- and accordingly the action \eqref{SHGO} becomes linear rather than quadratic in the momenta $P_a$.
The ``undeformed'' Hamiltonian action is then
\be
\begin{split}
S\Big|_{\lambda=0} = \int d^2\sigma \Big(&
 \dot{X}^i P_i- u X^{\prime i} P_i
 - \frac{e}{2} \left(
\delta^{ij}P_iP_j + \delta_{ij} X^{\prime i} X^{\prime j} 
\right) 
\\ & 
+ P_a ( \dot{X}^a - u X^{\prime a} - e \epsilon_b{}^a X^{\prime b} )
\Big)\,.
\end{split}
\label{SHGO_undeformed}
\ee
After integrating out the momenta $P_i$, for the coordinates $X^i$ alone we obtain the standard Polyakov action. In addition, we have the second term in \eqref{SHGO_undeformed} which is linear in the $P_a$.
The equations of motion of $P_a$ enforce that $X^0 \pm X^1$ are chiral/antichiral.
To see this, introduce a two-dimensional basis of vectors and covectors:
\be
x_a = \frac{1}{\sqrt{2}} 
\begin{pmatrix} 
1 \\ 1 
\end{pmatrix}
\,,\quad
\bar x_a = \frac{1}{\sqrt{2}} 
\begin{pmatrix} 
1 \\ - 1 
\end{pmatrix}\,,\quad
y^a = \frac{1}{\sqrt{2}} 
\begin{pmatrix} 
1 \\ 1 
\end{pmatrix}
\,,\quad
\bar y^a = \frac{1}{\sqrt{2}} 
\begin{pmatrix} 
1 \\ - 1 
\end{pmatrix}
\label{zerovectorsfirstlook}
\ee
such that $x_a y^a = 1$, $x_a \bar y^a =0$, $\bar x_a y^a = 0$ and $\bar x_a \bar y^a = 1$, and we have the relations
\be
x_a y^b + \bar x_a \bar y^b = \delta_a^b \,,\quad
x_a y^b - \bar x_a \bar y^b = \epsilon_a{}^b \,.
\ee
Define
\be
\bbeta \equiv y^a P_a \,,\quad \bbetabar \equiv \bar y^a P_a \,,\quad
\gamma \equiv x_a X^a \,,\quad \bar \gamma \equiv \bar x_a X^a \,.
\ee
In Lagrangian form, we then obtain the action
\be
\begin{split} 
S & = \int d^2\sigma - \frac{1}{2} \sqrt{- h} h^{\alpha \beta} \partial_\alpha X^i \partial^\alpha X^j  \delta_{ij}
 + \bbeta D_- \gamma 
 + \bbetabar D_+ \bar\gamma\,,
\end{split}
\label{SGO}
\ee
where the inverse worldsheet metric has components $h^{\tau\tau} =  -1/e^2$, $h^{\tau \sigma} = -u/e^2$, $h^{\sigma \sigma} = 1-u^2/e^2$, $\sqrt{-h} = e$, and $D_\pm \equiv \partial_\tau - u \partial_\sigma \pm e \partial_\sigma$. Conformal gauge corresponds to $e=1$, $u=0$.

From \eqref{SGO}, we see that the $(X^a,P_a)$ subsector now appears as a sum of chiral and antichiral $\bbeta \gamma$ systems. 
This is precisely the form of the non-relativistic string action of \cite{Gomis:2000bd}. 
The \emph{geometry} from which this subsector derives is in fact that of stringy Newton-Cartan \cite{Andringa:2012uz,Harmark:2017rpg, Harmark:2018cdl,Harmark:2019upf}, in which the vectors \eqref{zerovectorsfirstlook} play the role of singling out the preferred longitudinal time and space directions of the non-relativistic background probed by the string.
In section \ref{nonrieparam} below, we will review the interpretation of these vectors in terms of more general parametrisations of the generalised metric, describing non-Riemannian geometries including non-relativistic ones \cite{Lee:2013hma, Ko:2015rha,Morand:2017fnv}.

Now we describe how to view turning on $\lambda \neq 0$ as a deformation of the action \eqref{SHGO_undeformed}, first concentrating on how $\lambda$ appears in the generalised metric and then in the action itself.

\subsubsection*{The deformation as a TsT transformation of the generalised metric}

The $\lambda$ dependence of the generalised metric \eqref{GMGO} can be factorised out as follows:
\be
\mathcal{H}_{MN}= 
U_M{}^K(\lambda) U_N{}^L(\lambda) \mathcal{H}_{KL}(\lambda = 0) \,,
\label{Hppfac} 
\ee
where
\be
U_M{}^N(\lambda) = 
\begin{pmatrix}
\delta_a^b & 0 & 0 & 0 \\
0 & \delta_i^j & 0 & 0\\
\beta^{ab}(\lambda) & 0 & \delta^a_b & 0 \\
0 & 0 & 0 & \delta^i_j
\end{pmatrix} \,,\quad
\beta^{ab} =  \frac{\lambda}{2} \epsilon^{ab} \,,
\label{bivector}
\ee
and
\be
\mathcal{H}_{MN}(\lambda=0)=
\begin{pmatrix}
0  & 0 & \epsilon_a{}^b & 0\\
0 & \delta_{ij} & 0 & 0 \\
\epsilon_b{}^a &0 &   0 & 0 \\
0 & 0 & 0 &\delta^{ij} \\
\end{pmatrix} \,.
\label{GMGOnr}
\ee
The matrix $U_M{}^N(\lambda)$ in \eqref{bivector} is an element of $O(2,2;\mathbb{R}) \subset O(D,D;\mathbb{R})$ (i.e. it obeys $U_M{}^K(\lambda) U_N{}^L(\lambda) \eta_{KL} =\eta_{MN}$). This sort of $O(D,D)$ transformation we will call a \emph{bivector transformation}, referring to the antisymmetric quantity $\beta^{ab} = - \beta^{ba}$ appearing in \eqref{bivector}.

The matrix $\mathcal{H}_{MN}(\lambda=0)$ in \eqref{GMGOnr} is a \emph{non-Riemannian} or \emph{non-relativistic} generalised metric.
It can not correspond to the standard parametrisation of \eqref{GMRie}, as the bottom right block, which should correspond to $g^{\mu\nu}$, is not invertible.
This matrix \eqref{GMGOnr} describes the \emph{non-relativistic geometry} encoding the undeformed theory.

In general, bivector transformations can be factorised themselves as a T-duality on all directions for which the components $\beta^{\mu\nu}$ are non-zero (here just the longitudinal ones, so a T-duality on those two directions), followed by a constant shift of the $B$-field, followed by repetition of the same T-duality duality; alternatively they factor as a T-duality on a single direction, a shift of the coordinates, and a T-duality back on the same direction as before.
They are therefore one realisation of TsT transformations.
This TsT transformation of the non-relativistic background described by \eqref{GMGOnr} then amounts to a $T\bar T$ deformation.  This is the viewpoint arising from \cite{Baggio:2018gct,Araujo:2018rho, Frolov:2019nrr,Frolov:2019xzi,Sfondrini:2019smd}, so the relevance of $O(D,D)$ or TsT is of course not a new observation - what we want to focus on is the link to non-relativistic theories.

What is not clear from \eqref{GMGOnr} is how to make sense geometrically of the degeneracy of the bottom right block of the generalised metric. The general answer to this is provided by the classification of \emph{non-Riemannian parametrisations} of the generalised metric in \cite{Morand:2017fnv}, which we will review in section \ref{nonrieparam}.

\subsubsection*{The deformation as a current-current deformation of the action}

Now let's discuss how the action \eqref{SHGO_undeformed} is itself deformed when $\lambda \neq 0$.
The $\lambda$ dependence is really quite simple. We can write the action \eqref{SHGO} with $\lambda \neq 0$ as:
\be
S = S\big|_{\lambda=0} -\lambda \int d^2\sigma \frac{e}{2}  \eta^{ab}P_a P_b
= S\big|_{\lambda=0} + \lambda\int d^2\sigma e  \bbeta \bbetabar \,.
\ee
There is a nice description of the deformation in terms of a coupling to worldsheet currents (compare with the discussion in for instance \cite{Giveon:2017nie,Apolo:2018qpq,Araujo:2018rho,Apolo:2019zai} in particular regarding Wakimoto variables).
The action \eqref{SHGO} is invariant under translations $X^a \rightarrow X^a + \varepsilon^a$, implying a pair of (on-shell) conserved Noether currents
\be
J^\alpha_a = \begin{pmatrix}
P_a \\ 
- u P_a - e \epsilon_a{}^b P_b
\end{pmatrix} \,,\quad \partial_\alpha J^\alpha_a = 0 \,.
\ee
Then we have
\be
\frac{\partial S}{\partial \lambda} 
= 
\int d^2\sigma \frac{1}{4} \epsilon_{\alpha \beta} \epsilon^{ab} J^\alpha_a J^\beta_b \,.
\label{SJJ}
\ee
Equivalently, we could write these in terms of the chiral and antichiral currents associated to shifts in $\gamma$ and $\bar\gamma$.
In any case, the effect of the deformation is to recouple the $( \bbeta ,\gamma)$ and $(\bbetabar ,\gamma)$ subsectors via the introduction of a term involving $\bbeta \bbetabar$.
One can then integrate out $\bbeta$ and $\bbetabar$ to obtain the relativistic background with finite $\lambda$.

It is interesting to note that this sort of deformation has appeared in the calculation of the beta functionals of more general non-relativistic string actions \cite{Gomis:2019zyu, Gallegos:2019icg,Bergshoeff:2019pij}, arising at one-loop on the worldsheet. 
In that case, if one is interested in really restricting to non-relativistic target space geometries, one must ensure that the coefficient of the $\bbeta \bbetabar$ term vanishes identically in the non-relativistic background, as otherwise the background will again become relativistic.
A related discussion in the context of the equations of motions of double field theory i.e. the equations of motion of a theory in which the generalised metric is treated as the fundamental variable, can be found in \cite{Cho:2019ofr}. The question there concerns whether one should restrict to variations of $\cH_{MN}$ which preserve the non-relativistic parametrisation.

\subsubsection*{Upshot} 

The above discussion demonstrates how to view the $\lambda = 0$ worldsheet action as that of a string in a non-relativistic background.
The deformation with $\lambda \neq 0$ corresponds to deforming this action in a particular manner that corresponds to a certain TsT or bivector transformation of the geometry. 

\subsection{The relativistic and non-relativistic spectrum}
\label{spectra}

We will now present one further match between the non-relativistic and $T \bar T$ limits, namely the behaviour of the spectrum. 
Let's slightly generalise the background \eqref{initial}, following \cite{Ko:2015rha}, to:
\be
\begin{split}
\text{\underline{Gomis-Ooguri:}}\qquad
ds^2 &= \omega^2 ( - ( dX^0 )^2 + ( dX^1)^2 ) + \delta_{ij} dX^i dX^j \,, 
\\
B_{01}& = \omega^2 - \mu \,.
\end{split}
\label{GObg}
\ee
We have introduced a (finite) constant shift of the $B$-field, which can be incorporated easily into our previously analysis (for instance, in the Hamiltonian we just have $P_\mu \rightarrow\tilde P_\mu = P_\mu - B_{\mu\nu} X^{\prime \nu}$).

We now explicitly assume the direction $X^1$ is compact, with radius $R$, and that the string winds $w$ times around this direction. 
In fact, we should restrict to positive winding number: then physically what will happen is that the string states charged (positively) under the divergent $B$-field in \eqref{GObg} will survive in the $\omega^2\rightarrow\infty$ limit, with the divergent ``rest mass'' cancelling against the divergent contribution from the charge, and all other string states will decouple.

Standard string quantisation leads to the following spectrum of energies $E$:\footnote{Here we restore the effective string length squared, see appendix \ref{unitz}.}
\be
\frac{1}{\omega^2} \left( E + \frac{wR B_{01}}{\alpha_{\text{eff}}^\prime} \right)^2 = k^2 + \omega^2 \left( \frac{wR}{\alpha_{\text{eff}}^\prime} \right)^2 + \frac{1}{\omega^2} \left( \frac{n}{R} \right)^2 + \frac{2}{\alpha_{\text{eff}}^\prime} ( N_L + N_R - 2 ) \,,
\ee
where $k^2$ denotes the norm of the transverse spatial momenta.
Taking the square root gives \cite{Gomis:2000bd,Danielsson:2000gi}
\be
E - \frac{\mu w R}{\alpha_{\text{eff}}^\prime} 
= \frac{wR}{\alpha_{\text{eff}}^\prime} \omega^2 
\left(
 \sqrtbig{ 1
+ \frac{1}{\omega^2} \left( \frac{\alpha_{\text{eff}}^\prime}{wR} \right)^2 \left( k^2 + \frac{2}{\alpha_{\text{eff}}^\prime} (N+\tilde N - 2 ) \right)
+ \frac{1}{\omega^4} \left( \frac{\alpha_{\text{eff}}^\prime}{wR} \right)^2 \left( \frac{n}{R} \right)^2
}
-1
\right)\,,
\label{energyGOrel}
\ee
having chosen the sign such that the $\omega^2 \rightarrow \infty$ limit is well-defined.
This limit is:
\be
E(\omega^2 \rightarrow \infty) = \frac{\mu w R}{\alpha_{\text{eff}}^\prime} + \frac{\alpha_{\text{eff}}^\prime k^2}{2wR} + \frac{N_L + N_R -2}{wR} \,,
\label{energyGOnonrel}
\ee
which can be interpreted as a non-relativistic dispersion relation (energy equals momentum squared), assuming that $w>0$ so that the energy is positive.

To match with the $T \bar T$ literature, define:
\be
\lambda \equiv \frac{1}{\omega^2} 
\,,\quad 
r \equiv \frac{2wR}{\alpha_{\text{eff}}^\prime}
\,,\quad
\mathcal{E}( r,\lambda)\equiv E (r,\lambda) -\frac{1}{2}\mu r 
\,,\quad
\mathcal{E}(r,0) \equiv  \frac{1}{r} \left( k^2 + \frac{2}{\alpha_{\text{eff}}^\prime} (N+\tilde N - 2 )\right)\,,
\ee
so that
\be
\mathcal{E}(r,\lambda) = \frac{r}{2\lambda}
\left(
\sqrtbig{1 + \frac{4 \lambda}{r} \mathcal{E}(r,0)  
+ \frac{4\lambda^2}{r^2} \frac{n^2}{r^2} 
}
-1
\right) \,.
\ee
This is the known expression for the $T \bar T$ deformed spectrum obeying the \emph{inviscid Burger's equation} \cite{Zamolodchikov:2004ce,Smirnov:2016lqw, Cavaglia:2016oda,Jiang:2019hxb}.
\be
\partial_\lambda \mathcal{E}(r,\lambda) = \frac{1}{2} \partial_r \mathcal{E}(r,\lambda)^2 + \frac{1}{r} \frac{n^2}{r^2} \,.
\ee

\subsection{Lessons and morals}

We started with the background \eqref{initial} and compared two different ways of thinking about the limit in which $\lambda = \frac{1}{\omega^2} \rightarrow 0$.
One was as a non-relativistic limit \cite{Gomis:2000bd}, leading to the non-relativistic spectrum discussed in section \ref{spectra}, and (via the Hamiltonian formulation) the worldsheet action \eqref{SGO}.

The second point of view on this limit is that it coincides with a $T \bar T$ deformation ``in reverse'': we began with the parameter $\lambda \neq 0$ and then worked our way backwards to the point $\lambda = 0$. Then it is clear to see that the theory with $\lambda \neq 0$ has the same form as the $T\bar T$ deformation of the $\lambda = 0$ theory.

From the $T \bar T$ point of view, what one is interested in is the deformation of the two-dimensional theory obtained from the static gauge fixed Nambu-Goto action.
From a string theory perspective, we are interested in the corresponding deformation of the background geometry that contains much of the information about the deformation.
It has already been argued that the $T\bar T$ deformation corresponds to a TsT transformation. Here we see that we can even apply such transformations to non-relativistic geometries. We will continue the discussion of this sequence of transformations below.

We are suggesting therefore to view the non-relativistic limit as a ``reverse $T\bar T$ deformation''. Conversely, starting with the non-relativistic string, the $T\bar T$ deformation should make it relativistic. Effectively, one should think of the deformation parameter as turning on a finite speed of light.

The non-relativistic limit of \cite{Gomis:2000bd,Danielsson:2000gi} was partially inspired by work on the non-commutative open string limit \cite{Seiberg:2000ms, Gopakumar:2000na}. It might therefore be interesting to hunt for further links between these sort of string theory limits and $T \bar T$, in the context of open strings and D-branes. (Deformations linking Maxwell theory and DBI-style theories have been investigated in the $T \bar T$ literature in for example \cite{Conti:2018jho, Chang:2018dge, Brennan:2019azg}.)

\section{Non-relativistic duality and TsT}
\label{two}

We will now elaborate on the ideas that we have introduced by discussing in some more detail the more general setting.
We will introduce a set of general parametrisations of generalised metrics that allow the Hamiltonian formulation to describe more general non-relativistic backgrounds, and then connect our approach to a known recipe for $T \bar T$ as TsT.

\subsection{Non-Riemannian parametrisations}
\label{nonrieparam}

The systematic approach of \cite{Morand:2017fnv} (building on the examples found in \cite{Lee:2013hma, Ko:2015rha}) yields an elegant framework for dealing with generalised metrics for which a conventional spacetime interpretation is unavailable owing to the degeneracy of the $D \times D$ block which should correspond to the inverse spacetime metric.
A general classification has been provided based on the number of zero eigenvectors of this block and on the trace $\eta^{MN} \mathcal{H}_{MN}$ of the generalised metric, in terms of a pair of non-negative integers $(n,\bar n)$ such that the number of zero eigenvectors is $n+\bar n$ and the trace is $2(n-\bar n)$. 
The conventional case where we do have a spacetime metric and a $B$-field corresponds to $(0,0)$ while the generalised metric \eqref{GMGOnr} is an example of $(1,1)$. 
In general, $(1,1)$ appears to be relevant for stringy or torsional Newton-Cartan geometries.\footnote{Generic $(n,\bar n)$ parametrisations in fact correspond to changing the underlying coset to $O(D,D) / O(t+n,s+n) \times O(t+\bar n, s+\bar n)$, where $t+s+n+\bar n =D$. Here we focus only on the case $n=\bar n=1$ as this can be related to the standard description via $O(D,D)$ transformations.}

The $(1,1)$ parametrisation introduces two degenerate $D \times D$ matrices, $H^{\mu\nu}$ and $K_{\mu\nu}$, and is expressed in terms of a (particular) basis of zero vectors of these matrices, denoted $x_\mu, \bx_\mu$ and $y^\mu, \by^\mu$ with
\be
H^{\mu\nu} x_\nu = 0 = H^{\mu\nu} \bx_\nu\,,\quad
K_{\mu\nu} y^\nu = 0 = K_{\mu\nu} \by^\nu\,,\quad
x_\mu y^\mu = 1 = \bx_\mu \by^\mu\,,\quad
x_\mu \by^\mu = 0 = \bx_\mu y^\mu \,,
\label{HKdef}
\ee
plus a completeness relation
\be
H^{\mu \rho} H_{\rho \nu} + x_\nu y^\mu + \bx_\nu \by^\mu = \delta^\mu_\nu\,.
\label{HKcomplete}
\ee
In addition we can have a $B$-field, $B_{\mu\nu}$, and the generalised metric factorises as:
\be
\mathcal{H}_{MN}\Big|_{(1,1) \,\text{background}} = 
\begin{pmatrix}
\delta_\mu{}^\rho & B_{\mu \rho} \\
0 & \delta^\rho_\mu 
\end{pmatrix}
\begin{pmatrix} 
K_{\rho\sigma} & x_\rho y^\sigma - \bx_\rho \by^\sigma \\
y^\rho x_\sigma - \by^\rho \bx_\sigma & H^{\rho \sigma}
\end{pmatrix} 
\begin{pmatrix}
\delta^\sigma_\nu & 0 \\
- B_{\sigma \nu} & \delta^\nu_\sigma
\end{pmatrix} \,.
\ee
(Observe that the $B$-field factorisation is upper triangular whereas that of the bivector as in \eqref{bivector} is lower triangular.)
It is worth mentioning that the choice of the particular ingredients $(K,H,x,\bx, y,\by,B)$ is not unique, and in fact can be changed via a version of Galilean transformations \cite{Morand:2017fnv}.

The significance of the otherwise somewhat strange separation of the zero vectors into unbarred and barred becomes clear when we study the worldsheet action in such a background.
One way to do this is to simply use this generalised metric in the Hamiltonian constraint:
\be
\begin{split}
\mathcal{H}_e &  = \frac{1}{2} \begin{pmatrix} X^{\prime \mu} & \tilde P_\mu \end{pmatrix} 
\begin{pmatrix} 
K_{\mu\nu} & x_\mu y^\nu - \bx_\mu \by^\nu \\
y^\mu x_\nu - \by^\mu \bx_\nu & H^{\mu \nu}
\end{pmatrix} 
\begin{pmatrix} 
X^{\prime \nu} \\ \tilde P_\nu 
\end{pmatrix} 
\\ & 
= \frac{1}{2} K_{\mu\nu} X^{\prime \mu} X^{\prime \nu}
+ X^{\prime \mu} ( x_\mu y^\nu - \bx_\mu \by^\nu )\tilde P_\nu 
+ H^{\mu\nu}\tilde P_\mu \tilde P_\nu\,.
\end{split}
\ee
Here $\tilde P_\mu = P_\mu - B_{\mu\nu} X^{\prime \nu}$.
Ordinarily, the term quadratic in the momenta here would allow us to completely integrate them out of the action, and so return to the standard Polyakov Lagrangian (or to the Nambu-Goto Lagrangian after further integration out of $e$ and $u$). However, now $H^{\mu\nu}$ is degenerate, and to fully solve the equation of motion for $P_\mu$ we would need to invert this matrix. 
We can decompose $P_\mu$ (or $\tilde P_\mu$) using the completeness relation \eqref{HKcomplete}: $P_\mu = K_{\mu\rho}H^{\rho \nu} P_\nu + x_\mu y^\nu P_\nu + \bx_\mu \by^\nu P_\nu$.
The components proportional to $x_\mu$ and $\bx_\mu$, i.e. $y^\mu  P_\mu$ and $\by^\mu P_\mu$, appear linearly in the action and remain as independent fields, whose equations of motion turn out to enforce chirality or antichirality constraints. A straightforward calculation gives the action after integration out of $K_{\mu\rho}H^{\rho \nu} P_\nu$.
In conformal gauge ($e=1,u=0$) it is:
\be
\begin{split} 
S & = \int d^2\sigma \left( - \frac{1}{2} K_{\mu\nu} \partial_\alpha X^\mu \partial^\alpha X^\nu - \frac{1}{2} \epsilon^{\alpha \beta} B_{\mu\nu} \partial_\alpha X^\mu \partial_\beta X^\nu
 + \bbeta \,x_\mu \partial_- X^\mu 
 + \bbetabar \,\bx_\mu \partial_+ X^\mu\,,
 \right)
\end{split}
\label{covariantbosonic}
\ee
where we let $\bbeta \equiv y^\mu \tilde P_\mu$, $\bbetabar \equiv \by^\mu \tilde P_\mu$ and $\partial_\pm = \partial_\tau \pm \partial_\sigma$.
We see that our previous action \eqref{SGO} for the non-relativistic string obtained by the scaling limit is indeed of this form.

\subsection{TsT of non-relativistic geometry and worldsheet currents}

We now want to understand what happens to such a non-relativistic parametrisation, and to the action \eqref{covariantbosonic}, if we do a TsT transformation. 
How can we see the same sort of deformation that we encountered in our original example?
Let's first discuss this in terms of a deformation by worldsheet currents.
In general, invariance under $X^\mu \rightarrow X^\mu + \varepsilon^\mu$ in the Hamiltonian form of the string action gives the conserved current
\be
J_\mu^\alpha = \begin{pmatrix}
P_\mu \\ -u P_\mu - e \mathcal{H}_\mu{}^\nu P_\nu - e \mathcal{H}_{\mu\nu} X^{\prime \nu} 
\end{pmatrix}\,.
\ee
Suppose we have a background factorising as:
\be
\mathcal{H}_{MN} = U_M{}^K(\lambda) U_N{}^L(\lambda) \mathcal{H}_{KL}(\lambda=0)\,,\quad
U_M{}^N (\lambda) = \begin{pmatrix}
\delta_\mu^\nu & 0 \\
\lambda \beta^{\mu\nu} & \delta^{\mu}_\nu
\end{pmatrix}\,.
\label{assumefac}
\ee
where $\mathcal{H}_{MN}(\lambda=0)$ is a non-relativistic $(1,1)$ parametrisation.
Furthermore, assume that $\varepsilon^\mu = y^\mu \varepsilon$ and $\varepsilon^\mu = \by^\mu \bar\varepsilon$ both correspond to symmetries, i.e. both zero vector directions are (commuting) isometries and we have chosen coordinates such that the $y^\mu$ and $\by^\mu$ are constant.
This will remain a symmetry of the bivector transformed action (as the background remains independent of the coordinates corresponding to the $y^\mu$ and $\bar y^\mu$ directions).
We have two conserved currents: $J^\alpha \equiv y^\mu J_\mu^\alpha$ and $\bar J^\alpha \equiv \by^\mu J_\mu^\alpha$, for which
\be
\epsilon_{\alpha \beta} J^\alpha \bar J^\alpha 
 = 2 e y^{[\mu} \by^{ \nu]}  \left( P_\mu \mathcal{H}_{\nu}{}^\rho P_\rho 
- X^{\prime \nu} \mathcal{H}_{\rho \mu} P_\nu 
\right)\,.
\ee
Now, the off-diagonal block $\mathcal{H}_\nu{}^\rho$ can be seen to take the form $\mathcal{H}_{\nu}{}^\rho = x_\nu y^\rho - \bx_\nu y^\rho + $ terms involving the $B$-field of the non-relativistic parametrisation and involving the bivector.
We therefore have that:
\be
2 y^{[\mu} \by^{ \nu]}   P_\mu \mathcal{H}_{\nu}{}^\rho P_\rho
= - 2 y^\mu \by^\mu P_{\mu} P_{\nu} + \dots\,.
\ee
Hence this product of currents contains a term $y^\mu \by^\mu P_{\mu} P_{\nu} =\bbeta \bbetabar + \dots$ which is quadratic in the momenta, and recouples the formerly independent chiral and antichiral sectors.

Next, we can use the factorisation \eqref{assumefac} and the general expression for the action in Hamiltonian form \eqref{SHAM} to compute the dependence of the action on $\lambda$.
This is:
\be
\frac{\partial S}{\partial \lambda} 
= \int d^2\sigma e \beta^{\mu\nu} \left( P_\mu \mathcal{H}_{\nu}{}^\rho P_\rho 
- X^{\prime \nu} \mathcal{H}_{\rho \mu} P_\nu 
\right) \,.
\ee
If we construct our bivector using the zero vectors $y^\mu$ and $\by^\mu$, namely,
\be
\beta^{\mu\nu} = 2  y^{[\mu} \by^{\nu]} \,, 
\ee
then the deformation of the action takes the current-current form that we saw previously (equation \eqref{SJJ}) specialised to:
\be
\frac{\partial S}{\partial \lambda} 
 = \int d^2\sigma \epsilon_{\alpha \beta} J^\alpha \bar J^\beta \,.	
\ee
We therefore have a recipe to deform the non-relativistic geometry with longitudinal isometries via the $\lambda$-dependent bivector $\beta^{\mu\nu}(\lambda) = 2 \lambda y^{[\mu} \by^{\nu]}$.
This recovers and generalises the scaling limit deformation we discussed in the previous section.

There are multiple ways we could factorise this bivector transformation.
One would be to view it as resulting from T-duality in the $y^\mu$ direction, followed by a shift of the resulting dual direction by the $\bar y^\mu$ coordinate, $X^\mu\rightarrow Y^\mu = X^\mu - \lambda x_\nu \bar y^\mu X^\nu$, followed by T-duality back on the $y^\mu$ direction.
This gives the factorisation:
\be
\begin{pmatrix} 
\delta^\mu_\nu & 0 \\
2 \lambda y^{[\mu} \by^{\nu]} 
& \delta^\nu_\mu
\end{pmatrix} 
=
\begin{pmatrix}
\delta_\mu^\rho - x_\mu y^\rho & x_\mu x_\rho \\
y^\mu y^\rho & \delta^\mu_\rho - y^\mu x_\rho
\end{pmatrix} 
\begin{pmatrix}
\delta_\rho^\sigma + \lambda x_\rho \by^\sigma & 0 \\
0 & \delta_\sigma^\rho - \lambda \by^\rho x_\sigma 
\end{pmatrix} 
\begin{pmatrix}
\delta_\sigma^\nu - x_\sigma y^\nu & x_\sigma x_\nu \\
y^\sigma y^\nu & \delta^\sigma_\nu - y^\sigma x_\nu
\end{pmatrix} \,.
\label{biTsT}
\ee
It would be also natural to consider dualising along the directions picked out by $y^\mu \pm \by^\mu$ instead: this is what we will in fact describe next.

\subsection{The pp-wave example}

\subsubsection*{From Gomis-Ooguri to the pp-wave}

We have mentioned in the introduction the direct link between duality on null isometries and non-relativistic strings \cite{Harmark:2017rpg, Bergshoeff:2018yvt,Harmark:2018cdl,Harmark:2019upf}.
This seems initially to have nothing to do with the $\lambda \rightarrow 0$ limit we have been using.
Let's see what we can say about this, by rewriting the TsT transformation between the Gomis-Ooguri background \eqref{GObg} and its non-relativistic limit in terms of an explicit sequence of duality and shifts, similar to \eqref{biTsT}.
First recall that the background \eqref{GObg} was:
\be
\begin{split}
\text{\underline{Gomis-Ooguri:}}\qquad
ds^2 &= \frac{1}{\lambda} ( - ( dX^0 )^2 + ( dX^1)^2 ) + \delta_{ij} dX^i dX^j \,, 
\\
B_{01}& = \frac{1}{\lambda} - \mu \,.
\end{split}
\label{GObg1}
\ee
T-duality on the $X^1$ direction gives a background without a $B$-field:
\be
\begin{split}
\text{\underline{T of Gomis-Ooguri:}}\qquad
ds^2 &= 2 dX^0 ( d \tilde X^1 - \mu dX^0 ) + \lambda (d\tilde X^1 - \mu dX^0)^2 + \delta_{ij} dX^i dX^j \,.
\\
\end{split}
\label{GObgT}
\ee
In fact, the $\lambda \rightarrow 0$ limit of this background is well-defined. 
In this limit the (compact) direction $\tilde X^1$ becomes null. As pointed out in \cite{Gomis:2000bd}, this is the discrete lightcone quantisation (DLCQ) limit of string theory. The winding around the original direction $X^1$ becomes the null momentum.
Rather than take this limit though, we instead define shifted coordinates:
\be
Y^0 = X^0 + \frac{1}{2} \lambda (\tilde X^1 - \mu X^0) \,,\quad
\tilde Y^1 = \tilde X^1 - \mu X^0\,,
\ee
in terms of which we get the background
\be
\begin{split}
\text{\underline{Ts of Gomis-Ooguri:}}\qquad
ds^2 &= 2 dY^0 d\tilde Y^1 + \delta_{ij} dX^i dX^j \,.
\\
\end{split}
\label{GObgTs}
\ee
which directly manifests a null isometry in the $\tilde Y^1$ direction.
We can realise a null duality on this direction by acting directly on the generalised metric (see appendix \ref{buscher}).
This leads to the background we obtained as the $\lambda = 0$ limit of the original Gomis-Ooguri solution, i.e. we get the generalised metric $\mathcal{H}_{MN}(\lambda = 0)$ of equation \eqref{GMGOnr}.
In terms of the non-Riemannian $(1,1)$ parametrisation this is
\be
\begin{split}
K_{\mu\nu} = \begin{pmatrix} 0 & 0 & 0 \\ 0 & 0 & 0 \\ 0 & 0 & \delta_{ij} \end{pmatrix} \,,\quad
y^\mu = \frac{1}{\sqrt{2}} \begin{pmatrix} 1 \\ 1\\0 \end{pmatrix}\,,\quad 
\by^\mu = \frac{1}{\sqrt{2}} \begin{pmatrix} 1 \\ - 1 \\0\end{pmatrix} \,,
\\
H^{\mu\nu} = \begin{pmatrix} 0 & 0 & 0 \\ 0 & 0 & 0 \\ 0 & 0 & \delta^{ij} \end{pmatrix} \,,\quad
x_\mu = \frac{1}{\sqrt{2}} \begin{pmatrix} 1 \\ 1\\0 \end{pmatrix}\,,\quad 
\bx_\mu = \frac{1}{\sqrt{2}} \begin{pmatrix} 1 \\ - 1 \\0\end{pmatrix} \,.
\end{split}
\label{GOnrbits}
\ee
Suppose we had started with the background \eqref{GObgTs}, or more generally any background with a null isometry, generated by a null Killing vector which let us denote by $\frac{\partial}{\partial U}$.
Instead of appealing to unfamiliar generalised metrics, how could we take a duality in the null direction?
If there is a second (commuting) isometry present, generated by $\frac{\partial}{\partial X}$, one solution would be to define a duality along the isometry generated by a linear combination of the two Killing vectors, specifically by $\frac{\partial}{\partial U} + \frac{\lambda}{2} \frac{\partial}{\partial X}$ (assuming this is non-null), and then take the limit $\lambda \rightarrow 0$ at the end. This is the same as defining shifted coordinates $\tilde X = X - \frac{\lambda}{2} U$, $\tilde U = U$, and dualising along the $\tilde U$ isometry, $\frac{\partial}{\partial \tilde U} =  \frac{\partial}{\partial U} + \frac{\lambda}{2} \frac{\partial}{\partial X}$.
Applying this procedure to \eqref{GObgTs} generates the background \eqref{GObgT}, and then \eqref{GObg}, in which the initial difficulty in considering a null duality shows up again as the singular behaviour as $\lambda \rightarrow 0$.

\begin{figure}[h]
\centering
\begin{tikzpicture}

\node [text width=4cm, align=center,draw, thick,white,rounded corners=5pt] (nR) at (-5,-2) {\color{black} Non-relativistic \eqref{GOnrbits} \\ ($Y^0, Y^1$)};

\node [text width=4cm, align=center,draw, thick,white,rounded corners=5pt] (pp) at (-5,2) {\color{black}  pp-wave \eqref{GObgTs}\\ ($Y^0,\tilde Y^1$)};

\node [text width=4cm, align=center,draw, thick,white,rounded corners=5pt] (ppshift) at (5,2) {\color{black} shifted pp-wave \eqref{GObgT}\\ ($X^0,\tilde X^1$)};

\node [text width=4cm, align=center,draw, thick,dashed,white,rounded corners=5pt] (ppdef) at (5,-2) {\color{black}  GO (relativistic) \eqref{GObg1} \\ ($X^0,X^1$)};

\draw [thick,<->] (nR) -- (pp) node [midway, left] {T: $Y^1 \leftrightarrow \tilde Y^1$ (null)};

\draw [thick,<->] (pp) -- (ppshift) node [midway, above] {s: $(Y^0,\tilde Y^1) \leftrightarrow (X^0, \tilde X^1)$};

\draw [thick,<->] (nR) to [out=0,in=180,->]   node [midway,above] {TsT} (ppdef);

\draw [thick,<->] (ppshift) -- (ppdef) node [midway, right] {T: $\tilde X^1 \leftrightarrow X^1$};

\end{tikzpicture}
\caption{Duality between our example non-relativistic and relativistic backgrounds	}
\label{fig:example}
\end{figure}
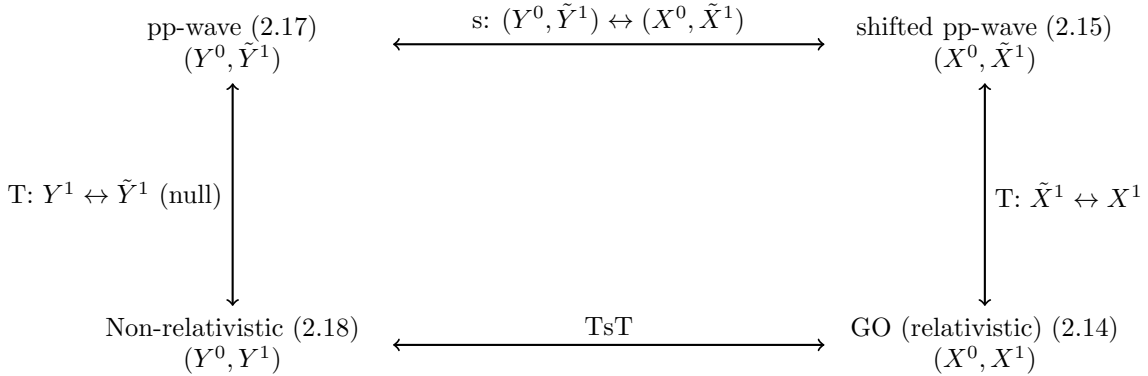

This chain of transformations is illustrated in figure \ref{fig:example}.

\subsubsection*{A general recipe for $T\bar T$ as TsT}

\begin{figure}[h]
\centering
\begin{tikzpicture}

\node [text width=4.25cm, align=center,draw, thick,white,rounded corners=5pt] (nR) at (-5,-2) {\color{black} Undeformed geometry \\ ($Y^0, Y^1$)};

\node [text width=4.25cm, align=center,draw, thick,white,rounded corners=5pt] (pp) at (-5,2) {\color{black} Dual undeformed geometry \\ ($Y^0,\tilde Y^1$)};

\node [text width=4.25cm, align=center,draw, thick,white,rounded corners=5pt] (ppshift) at (5,2) {\color{black} Deformed dual geometry\\ ($X^0,\tilde X^1$)};

\node [text width=4.25cm, align=center,draw, thick,dashed,white,rounded corners=5pt] (ppdef) at (5,-2) {\color{black}  Deformed geometry \\ ($X^0,X^1$)};

\draw [thick,<->] (nR) -- (pp) node [midway, left] {T: $Y^1 \leftrightarrow \tilde Y^1$};

\draw [thick,<->] (pp) -- (ppshift) node [midway, above] {s: $(Y^0,\tilde Y^1) \leftrightarrow (X^0, \tilde X^1)$};

\draw [thick,<->] (nR) to [out=0,in=180,->]   node [midway,above] {TsT} (ppdef);

\draw [thick,<->] (ppshift) -- (ppdef) node [midway, right] {T: $\tilde X^1 \leftrightarrow X^1$};

\node [text width=5cm] (static) at (6,-3.5) {Static gauge\\ $\Rightarrow$ $T\bar T$ deformed theory};
\draw [->] (ppdef) to [out=270,in=90] (static);

\node [text width=5cm] (staticundef) at (-6,-3.5) {Static gauge\\ $\Rightarrow$ undeformed theory};
\draw [->] (nR) to [out=270,in=90] (staticundef);

\node [text width=5cm] (ulcg) at (6,3.5) {ULCG \\$\Rightarrow$ $T\bar T$ deformed theory};
\draw [->] (ppshift) to [out=90,in=270] (ulcg);

\node [text width=5cm] (ulcgundef) at (-6,3.5) {ULCG \\$\Rightarrow$ undeformed theory};
\draw [->] (pp) to [out=90,in=270] (ulcgundef);

\end{tikzpicture}
\caption{A recipe for $T \bar T$ as TsT following \cite{Baggio:2018gct, Frolov:2019nrr,Frolov:2019xzi,Sfondrini:2019smd}}
\label{fig:recipe}
\end{figure}
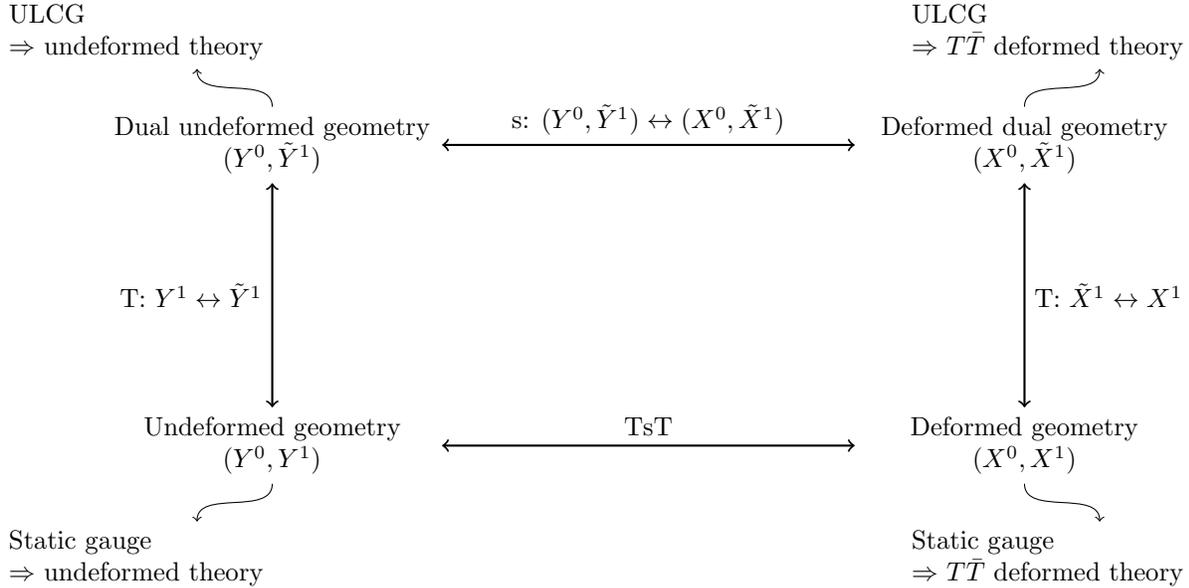

This picture can be compared with the general procedure advocated in \cite{Baggio:2018gct, Frolov:2019nrr,Frolov:2019xzi,Sfondrini:2019smd}, see figure \ref{fig:recipe}.
Let's outline (a somewhat simplified version of) the procedure of \cite{Sfondrini:2019smd}.
The idea is to work within the arena of string sigma models in backgrounds with two isometries (one timelike).
The two isometric directions, which we again call longitudinal, are to be viewed as the ``extra'' two coordinates which appear fixed to static gauge in the $T\bar T$ deformation of some two-dimensional theory. The question is then how to realise such a deformation directly in stringy language.
This is achieved by the recipe depicted diagrammatically in figure \ref{fig:recipe}.

One starts with the undeformed geometry in the bottom left and T-dualises on the $Y^1$ coordinate.
Then one performs a shift in the dual ``undeformed'' geometry, of the form $Y^0=X^0+\frac{1}{2}\lambda \tilde X^1$, $\tilde Y^1 = \tilde X^1$ (where $\tilde Y^1$ is dual to $Y^1$).
After shifting, and thereby introducing the parameter $\lambda$, one T-dualises on $\tilde X^1$ to reach the deformed geometry in the bottom right hand corner of figure \ref{fig:recipe}.
 
After arriving at the deformed geometry in the bottom right hand corner of figure \ref{fig:recipe}, fixing static gauge (in the Hamiltonian form of the action, say) allows one to recover the $T \bar T$ deformed theory which for $\lambda = 0$ would be encoded by the original undeformed geometry.\footnote{This sequence gives solely a $T \bar T$ deformation. One can generalise to include other shifts which lead to different deformations, inlcuding in the situation where there is a third isometry present allowing for $JT$-type deformations \cite{Frolov:2019xzi}, which we do not discuss here. What is important is that although the shift looks locally like a diffeomorphism, globally it is generates a different geometry and hence can be interpreted as a deformation rather than a gauge transformation.}
Alternatively, instead of going to static gauge, one could consider just the dual deformed geometry with coordinates $X^0$ and $\tilde X^1$. In uniform light cone gauge (ULCG) this reproduces the same underlying $T \bar T$ deformed theory (this gauge is $X^0 \sim \tau$ and $p_1 \sim$ constant; the momenta $p_1$ is dual to the winding of $X^1$. This uniform light cone gauge approach is the principal focus of \cite{Frolov:2019nrr,Frolov:2019xzi}).

One example of this sort considered in detail in \cite{Sfondrini:2019smd} is that of a pp-wave.
Using our own conventions for the names of the coordinates\footnote{My priorities are backwards to those of \cite{Sfondrini:2019smd} (if not also in general), in that I have been interested in starting with deformed geometries and working my way back to undeformed geometries: as a result, the coordinates I call $X$ are those \cite{Sfondrini:2019smd} call $Y$ and vice versa. I will also use indices $0,1$ instead of $+,-$, but these are just labels and should not be construed as necessarily attaching any meaning regarding the ``lightcone'' nature of coordinates, as will be clear from the explicit metrics.}, we choose to write the metric as follows:
\be
\text{\underline{pp-wave:}}\qquad
ds^2 = 2 dY^0 d\tilde Y^1 - 2 V(X^i) dY^0 dY^0 + dX^i dX^j \delta_{ij} \,,
\label{pp}
\ee
(where we chose the numerical factors for later convenience; they are related to those of \cite{Sfondrini:2019smd} by a rescaling of the coordinates). 
The function $V(X^i)$ is either constant or quadratic.
A priori, it appears that this cannot be dualised on $\tilde Y^1$, as this is a null isometry. 
Nevertheless, one can shift the coordinates as $X^0 = Y^0 - \frac{1}{2} \lambda \tilde Y^1$, $\tilde X^1 = \tilde Y^1$, before T-dualising on $\tilde X^1$ to arrive at the background:\footnote{The renaming and rescaling of the coordinates relative to \cite{Sfondrini:2019smd} is $X^0_{\text{here}} = \frac{1}{\sqrt{2}} Y^+_{\text{there}}$ and $X^1_{\text{here}} \equiv \frac{1}{2\sqrt{2}} \tilde Y^-_{\text{there}}$)}
\be
\begin{split}
\text{\underline{Deformed pp-wave:}}\qquad
ds^2 & =  \frac{2}{\lambda( 2-\lambda V)} \left( - dX^0 dX^0 + d X^1 d X^1\right)  + dX^i dX^j \delta_{ij} \,,\\
B_{01} & = \frac{2}{\lambda} \frac{1 - \lambda V}{2- \lambda V} \,.
\end{split} 
\label{pplambda}
\ee
This geometry is then an ``sT'' transformation of the pp-wave \eqref{pp}.
Naively, we cannot view it as the ``TsT'' transformation of the geometry which would be T-dual to the pp-wave along the null isometry direction $\tilde Y^1$.
As explained in \cite{Sfondrini:2019smd}, this manifests itself in the fact that the background \eqref{pplambda} is singular for $\lambda \rightarrow 0$. 
However, the Hamiltonian is well-behaved in this limit, which means one can still gauge fix and find a resulting Hamiltonian model for the coordinates $X^i$ and their momenta, which can be interpreted as a $T\bar T$ deformation of the pp-wave Hamiltonian, $H = \frac{1}{2} P_i P_j \delta^{ij} + \frac{1}{2} X^{\prime i} X^{\prime j} \delta_{ij} + V(X^i)$.

What we can now is interpret $\lambda \rightarrow 0$ limit as corresponding to a non-relativistic geometry. 
Indeed, if $V=0$ the background is exactly \eqref{GObg1} with $\mu=0$, i.e. our initial background \eqref{initial}.

\subsubsection*{Non-Riemannian parametrisation describing the deformed pp-wave}

We again denote the two (longitudinal) coordinates of the background \eqref{pplambda} (those that we have been acting on with T-dualities and shifts) as $X^a = (X^0, X^1)$, such that altogether $X^\mu = (X^0,X^1,X^i)$.
The background \eqref{pplambda} is then described by the following generalised metric
\be
\text{\underline{Deformed pp-wave:}}\quad
\mathcal{H}_{MN}  = 
\begin{pmatrix}
2V \eta_{ab}  & 0 & (1-\lambda V) \epsilon_a{}^b & 0\\
0 & \delta_{ij} & 0 & 0 \\
(1-\lambda V) \epsilon_b{}^a &0 &   \frac{\lambda}{2} (2 - \lambda V) \eta^{ab} & 0 \\
0 & 0 & 0 &\delta^{ij} \\
\end{pmatrix} \,.
\label{Hpp}
\ee
This is non-Riemannian for $\lambda \rightarrow 0$, and factorises as in \eqref{Hppfac} terms of the very same bivector transformation \eqref{bivector} as before, except now
\be
\mathcal{H}_{MN}(\lambda=0)=
\begin{pmatrix}
2V \eta_{ab}  & 0 & \epsilon_a{}^b & 0\\
0 & \delta_{ij} & 0 & 0 \\
\epsilon_b{}^a &0 &   0 & 0 \\
0 & 0 & 0 &\delta^{ij} \\
\end{pmatrix} \,.
\label{Hppnr}
\ee
A convenient decomposition for this generalised metric \eqref{Hppnr} is provided by the same parametrisation \eqref{GOnrbits} we used for the original non-relativistic background  with in addition the potential $V(X^i)$ encoded in the $B$-field:
\be
B_{\mu\nu} = \begin{pmatrix} 0 & -V & 0 \\ V & 0 & 0 \\ 0 & 0 & 0 \end{pmatrix} \,.
\ee
With this information, the worldsheet action \eqref{covariantbosonic} describes the $\lambda = 0$ non-relativistic geometry of the deformed background \eqref{pplambda}.
Turning on the deformation ($\lambda \neq 0$) immediately returns us to a relativistic description.

Let us mention that aficionados of torsional Newton-Cartan should identify the coordinate here called $X^1$ (or previously $Y^1$ in the $\lambda \rightarrow 0$ limit) with the extra worldsheet coordinate which is dual to a null isometry direction of a relativistic background - this is the coordinate called $\eta$ in \cite{Harmark:2019upf}. Then if $X^I =(X^0,X^i)$ label the coordinates of the Newton-Cartan geometry, the Newton-Cartan clock form is $\tau_I=\delta_I^0$, its dual vector is $v^I = - \delta^I_0$, and the $U(1)$ gauge field is $m_I = V \delta_I^0$, while the degenerate matrix $h_{IJ}$ has non-zero components $h_{ij} = \delta_{ij}$, similarly $h^{IJ}$ has non-zero components $h^{ij} = \delta^{ij}$.

\subsubsection*{Gauge fixing}

For completeness, let's exhibit how the gauge fixing procedure of \cite{Sfondrini:2019smd} works, in order to relate back to the actual $T\bar T$ deformation picture.
We are supposed to fix static gauge and interpret the resulting theory as a $T\bar T$ deformation. 
We define static gauge by $X^0 = \tau$ and $X^1 =  \sigma$. 
The gauge fixed action after solving the constraints takes the form:
\be
S = \int d^2 \sigma \dot{X}^i P_i - H \,,\quad
H \equiv - P_0 \,.
\label{Sgf}
\ee
We solve the first constraint, $\mathcal{H}_u=0$, for the momentum $P_1$:
\be
P_1 = - 
X^{\prime i}P_i \,,
\ee
and the second constraint, $\mathcal{H}_e = 0$, for the momentum $P_0$:
\be
\begin{split}
- P_0 = & \frac{2}{\lambda (2 -\lambda V)} \Bigg( 
\sqrtbig{
1+ \lambda H_\perp 
+ \lambda^2 \left( 
P_1^2 - \frac{1}{2} V H_\perp
\right) 
- {\lambda^3}V P_1^2 
+ \frac{\lambda^4}{4} V^2 P_1^2 
} - 1 + \lambda V
\Bigg)\,, 
\end{split}
\label{gfHam}
\ee
identifying
\be
H_{\perp} \equiv \delta^{ij} P_i P_j + \delta_{ij} X^{\prime i} X^{\prime j} \,.
\ee
This agrees with the expression in \cite{Sfondrini:2019smd} up to the rescalings of the coordinates (and a possible minus sign typo).
We chose the sign before the square root in order that the non-relativistic $\lambda \rightarrow 0$ limit is finite. 
In this limit we have
\be
- P_0 =   \frac{1}{2} (\delta^{ij} P_i P_j + \delta_{ij} X^{\prime i} X^{\prime j}) + V + O(\lambda)\,.
\ee

\subsection{The ``negative string'' example}
\label{negativeF1}

We are now going to write down an example of an apparently singular SUGRA solution where the ``flow'' from non-relativistic to relativistic is realised geometrically as we move in the background.
This is a background for which the author has a certain problematic fondness \cite{Blair:2016xnn} and is realised by performing dualities on the fundamental string solution along its isometries in both the spatial worldvolume direction and the time direction. This example has provided a useful inspiration for the study of non-Riemannian backgrounds in string theory \cite{Lee:2013hma, Park:2015bza,Ko:2015rha} and can also be viewed as a negative tension brane \cite{Dijkgraaf:2016lym}.

The metric and $B$-field of this solution are:
\be
\begin{split}
\text{\underline{Negative F1:}}\qquad
ds^2 &= \frac{1}{f_-(r)} ( - ( dX^0 )^2 + ( dX^1)^2 ) + \delta_{ij} dX^i dX^j \,, 
\\
B_{01}& = \frac{1}{f_-(r)} - 1 \,,
\end{split}
\label{negF1}
\ee
where we define harmonic functions
\be
f_\pm (r) = 1 \pm \left( \frac{r_0}{r} \right)^6 \,,\quad r \equiv \sqrt{\delta_{ij} X^i X^j } \,.
\ee
The standard fundamental string solution would have the form of \eqref{negF1} but with $f_+$ in place of $f_-$.
There is a naked singularity at $r=r_0$ (matching the SUGRA solution with a fundamental string source for the normal string solution gives $r_0\equiv 2\pi \sqrt{\alpha^{\prime}} / ( 6 \mathrm{Vol}(S^7))^{1/6}$). 
At this singularity, $f_- \rightarrow 0$. 
The form of the metric and $B$-field is similar to the Gomis-Ooguri background \eqref{GObg1}, or our initial spacetime \eqref{initial}, which is unsurprising as the form of these backgrounds evidently mirror that of the fundamental string solution.
In place of the parameter $\lambda$ that we tuned ``manually'' we now have the function $f_-(r)$ which depends on our position in the transverse space to the string.
In static gauge, the Nambu-Goto action for the string in the background \eqref{negF1} is
\be
\begin{split}
S_{NG}
= \int d^2 \sigma \frac{1}{f_-(r)} \left(
1 - f_-(r) - \sqrtbig{ 1 - f_-(r) \delta_{ij} ( \dot{X}^i \dot{X}^j - X^{\prime i} X^{\prime j} ) 
 - f_-(r)^2 \det (\partial_\alpha X^i \partial_\beta X^j \delta_{ij} )
 }
\right)
\end{split}
\ee
and, seeing as we now know very well how this behaves for $f_-(r) \rightarrow 0$, is non-singular at $r=r_0$. This is similar to the analysis of \cite{Dijkgraaf:2016lym} that showed that apparently singular negative brane solutions can be safely probed by other (mutually BPS) branes.

The generalised metric for this background factorises as 
\be
\mathcal{H}_{MN}(r) = (U_{\beta})_M{}^K(r)(U_{B})_K{}^L(r)  (U_{\beta})_N{}^P(r)(U_{B})_P{}^Q(r) \mathcal{H}_{LQ}(r_0) \,,
\ee
where at $r=r_0$ we have a non-Riemannian parametrisation (writing only the part of the generalised metric involving the longitudinal directions $X^a = (X^0, X^1)$):
\be
\mathcal{H}_{MN}(r_0) = \begin{pmatrix} 
2 \eta_{ab} & \epsilon_a{}^b \\
 \epsilon_b{}^a & 0 
\end{pmatrix} 
\ee
and we have factorised out both a $B$-field contribution:
\be
(U_B)_M{}^N (r) = \begin{pmatrix}
\delta_a^b & \frac{1}{2} f_-(r) \epsilon_{ab} \\
0 & \delta_b^a
\end{pmatrix} \,,
\label{negF1be}
\ee
and a bivector contribution:
\be
(U_\beta)_M{}^N(r) =
\begin{pmatrix}
\delta_a^b & 0 \\
\frac{f_-(r)}{f_+(r)} \epsilon^{ab} & \delta_b^a
\end{pmatrix} \,.
\label{negF1bi}
\ee
(Here $\epsilon^{01} = -\epsilon_{01} = -1$.)
Both \eqref{negF1be} and \eqref{negF1bi} vanish at $r=r_0$. 
The bivector factorisation shows that we can interpret the deformation away from the non-relativistic locus at $r=r_0$ as a sort of TsT transformation, albeit now an $r$-dependent one.
This example can be related back to the pp-wave solution which is T-dual to the fundamental string. In the pp-wave, the Killing vector associated to the isometry in the time direction is timelike for $r>r_0$ and null at $r=r_0$, leading to the degeneracy at that point in the dual solution.

Starting at $r=\infty$ and moving towards the apparent singularity at $r=r_0$ corresponds to moving from $f_-(r) = 1$ to $f_-(r)=0$.
We can continue to probe values of $r<r_0$ (from \cite{Dijkgraaf:2016lym} we can view this as a ``bubble'' surrounding the position of the brane, in which spacetime signature flips and physics should be described by an exotic variant of string theory), which corresponds to a region where $f_-(r)$ is negative, with the position of the string itself corresponding to $f_-(r)\rightarrow -\infty$.
This perhaps correspond to a ``wrong sign'' of the $T \bar T$ parameter, and may be worth further attention.
(Meanwhile we should not forget the dilaton, which is $e^{\phi} = | f_-(r)|^{-1/2}$, blowing up at $r=r_0$ and goes to zero at $r=0$. One could make sense of the strong coupling behaviour either using S-duality or by uplifting to a smooth configuration in 11 dimensions \cite{Dijkgraaf:2016lym}.)

\subsection{From non-relativistic to ultra-relativistic}
\label{ultra}

We interpreted the undeformed backgrounds with $\lambda = 0$ as corresponding to non-relativistic geometries.
For finite $\lambda$, we recovered standard relativistic backgrounds.
If we follow our intuition that $\lambda$ is related to the inverse speed of light squared, then $\lambda \rightarrow \infty$ should send the speed of light to zero.
This is an \emph{ultra}-relativistic limit, and is whimsically known as the Carrollian limit \cite{LevyLeblondCarroll}.

Some physical intuition for this limit can be obtained by considering the slopes of lightcones, thinking of the basic equation $t = \pm \frac{1}{c} x$). For $c \rightarrow \infty$, the lightcones at any point expands to fill the whole future region as the speed of signal propagation becomes infinite. For $c \rightarrow 0$ on the other hand the slope of the lightcones becomes steeper (moving towards the time axis), so that eventually the lightcones shrink and coincide with the past and future time axes. This means that free particles cannot in fact move spatially (as they still cannot travel outside their lightcone), and so are frozen in place (see for instance \cite{Duval:2014uoa}). 

In \cite{Morand:2017fnv} the behaviour of both particles and strings in non-Riemannian geometries were studied, encompassing both Newton-Cartan or Gomis-Ooguri non-relativistic limits and Carollian ultra-relativistic limits. For each zero vector of the degenerate block $H^{\mu\nu}$ in the generalised metric, we find a momentum that we cannot integrate out, leading to an equation of motion for the corresponding target space coordinate which implies for a particle that the coordinates in the directions picked out by the zero vectors are constant, $x_\mu \dot{X}^\mu=0$, while for a string the coordinates (as we have seen) become chiral or antichiral, $x_\mu (\dot{X}^\mu - X^{\prime \mu}) = 0$, $\bar x_\mu ( \dot{X}^\mu + X^{\prime \mu}) = 0$.

What can we say about this limit? 
First, consider the expression for the spectrum obtained in \ref{spectra}.
The limit as $\lambda \rightarrow \infty$ of \eqref{energyGOrel} is
\be
E(\lambda \rightarrow \infty) =\left| \frac{n}{R} \right| \,.
\label{spectrumGOCarroll}
\ee
In this limit, the energy of the string is given by its momentum in the longitudinal direction. 
There is no energy contribution coming from the transverse directions.

Next, let's consider the action in Hamiltonian form coming from static gauge fixing, given by \eqref{Sgf} (with or without $V=0$). 
The Hamiltonian there as $\lambda \rightarrow \infty$ is:
\be
H = \sqrt{ P_1^2} = \sqrt{ ( X^{\prime i} P_i )^2 }\,.
\ee
This agrees with \eqref{spectrumGOCarroll}, identifying $P_1 = \frac{n}{R}$. 
Now, the other Hamiltonian constraint gave us what is really the level-matching condition $P_1 = - X^{\prime i}P_i$.
If $n>0$, then $P_1 >0$ and $X^{\prime i} P_i$ is negative.
Then the Hamiltonian form of the action in the limit is:
\be
S = \int d^2\sigma P_i ( \dot{X}^i  +  X^{\prime i} ) \,,
\label{Carroll1}
\ee 
conversely if $n<0$ then $P_1<0$ and $X^{\prime i} P_i > 0$, so we get the action
\be
S = \int d^2\sigma P_i ( \dot{X}^i  -  X^{\prime i} ) \,.
\label{Carroll2}
\ee
These correspond to chiral or antichiral $\bbeta \gamma$ systems as we would expected to obtain associated to the coordinates in  the zero vector directions in a non-relativistic parametrisation.
This is what we should expect in this situation \cite{Morand:2017fnv} if we can really interpret this limit as giving us a Carrollian geometry (which is a $(D-2,0)$ parametrisation of the generalised metric).
A possible interpretation of this picture is the following: in this ultra-relativistic limit the string becomes chiral in the subsector where it is moving one way around the longitudinal spatial circle and antichiral in the subsector where it is moving the other way around this direction.

For a single transverse scalar, the above limit was considered in \cite{Townsend:2019koy} and used to provide a description of a chiral boson starting with the gauge fixed Nambu-Goto action; in \cite{Chakrabarti:2020pxr} this was reinterpreted in the $T\bar T$ context and the presence of chiral and antichiral sectors also suggested.

\subsection{Lessons and morals}

In this section we have discussed some of the general features of the description of non-relativistic geometries in string theory using so-called non-Riemannian parametrisations of the generalised metric that appears in the Hamiltonian. 
We made contact with results about $T \bar T$ as a TsT transformation and added the non-relativistic interpretation of the bottom left node of the diagram shown in figure \ref{fig:recipe}. We explained how to realise this TsT transformation in terms of the zero vectors characterising the part of the geometry to which the longitudinal directions of the string couple.  
In section \ref{negativeF1} we wrote down an example of a background where the ``flow'' or ``deformation'' was from an apparently singular non-relativistic locus at $r=r_0$ to a well-behaved relativistic background for $r>r_0$. This may be an isolated curiousity, or perhaps a guide for realising these ideas in an intrinsically geometric fashion in other backgrounds.
(This raises the more general idea of reanalysing the string action in non-geometric T-folds patched by bivector transformations in terms of worldsheet deformations.)

As $T \bar T$ deformations preserve integrability, one might wonder what one can say about non-relativistic limit of the AdS${}_5 \times S^5$ superstring \cite{Gomis:2005pg
} (integrability of Newton-Cartan strings has been recently studied in \cite{Roychowdhury:2019vzh}).
For this the Green-Schwarz doubled string of \cite{Park:2016sbw} could be a good place to start - this paper already considered the Gomis-Ooguri non-relativistic string as an example (an RNS doubled string was applied to non-relativistic backgrounds in \cite{Blair:2019qwi}). The results of for instance \cite{Baggio:2018rpv,Frolov:2019nrr} on the $T \bar T$ side may be of use here.

\section{M2 brane deformations and non-relativistic U-duality}
\label{mdef}

Non-relativistic limits can be taken not just for particles and strings, but also for general branes \cite{Gomis:2000bd, Gomis:2004pw}.
This suggests an experimental approach to understanding possible higher-dimensional generalisations of the $T \bar T$ deformation (discussed for instance in \cite{Cardy:2018sdv,Taylor:2018xcy,Bonelli:2018kik}): start with a higher-dimensional Nambu-Goto style action and try to make sense of the non-relativistic limit in some generalised $T \bar T$ sense.
We will therefore examine in this section what the analogous non-relativistic limit looks like for M2 branes. 

\subsection{Limits of the Dirac-Nambu-Goto action}
\label{DNG}

We will restrict our attention to the M2 brane in 11 dimensions.
Let $\sigma^{A} = (\tau,\sigma^1,\sigma^1)$ denote the worldvolume coordinates, and $\epsilon^{ABC}$ the alternating symbol with $\epsilon^{\tau12} = -1$. The bosonic action is
\be
S_{\text{M2}} = 
- 
\int d^3 \sigma \left( \sqrt{-\det g_{AB}} + \frac{1}{6} \epsilon^{ABC} C_{ABC}\right)
\,,
\label{SM2}
\ee
featuring the pullbacks of the 11-dimensional metric, $g_{AB} = \partial_A X^\mu \partial_B X^\nu g_{\mu\nu}$, and the three-form, $C_{ABC} = \partial_A X^\mu \partial_B X^\nu \partial_C X^\rho C_{\mu\nu\rho}$.
An appropriate flat space background for the limit we want to take is:
\be
\begin{split}
ds^2 & = \omega^{-2/3} \left( \omega^2 \eta_{ab} dX^a dX^b + \delta_{ij} dX^i dX^j \right)\,,\\
C_{012} & = \omega^2\,.
\end{split} 
\label{M2GO}
\ee
The overall factor of $\omega^{-2/3}$, and the choice of the three-form, ensures that we obtain a finite limit from the Nambu-Goto action in the $\omega^2 \rightarrow \infty$ limit (recall $\omega^2$ is dimensionless).
Alternatively, we could omit this overall scaling of the metric, as in \cite{Gomis:2004pw}, in which case we would need to also rescale the membrane tension to get a finite action in the limit. 
We will continue to use the background \eqref{M2GO} as in \cite{Berman:2019izh} and not scale the tension (which we continue to suppress from our expressions).
Observe that the powers of $\omega^{-2}$ match the appearance of the harmonic function in the M2 brane SUGRA solution.

Static gauge for the membrane is: $X^0 = \tau, X^1 = \sigma^1, X^2 =\sigma^2$.
Let $\eta_{AB}$ denote the Minkowski metric of the three-dimensional theory.
The following compressed notation is useful:
\be
X_{AB} \equiv \partial_A X^i \partial_B X^j \delta_{ij} \,,\quad
\mathrm{tr}\,X \equiv \eta^{AB} X_{AB} \,,\quad
\mathrm{tr}\, ( \eta^{-1} X \eta^{-1} X ) \equiv
\eta^{AB} X_{BC} \eta^{CD} X_{DA}\,.
\ee
Using this, the action \eqref{SM2} becomes:
\be
S_{\text{M2}} \Big|_{\text{static gauge}} 
=
\int d^3 \sigma
\frac{1}{\lambda}
\left(
1 - \sqrtbig{1 + \lambda \mathrm{tr}\,X + \frac{1}{2} \lambda^2 \left( (\mathrm{tr} \, X )^2 - \mathrm{tr}\, ( \eta^{-1} X \eta^{-1} X )\right) - \lambda^3 \det X}
\right) \,.
\label{SM2static}
\ee
The expansion for $\lambda \rightarrow 0$ gives
\be
S_{\text{M2}} \Big|_{\text{static gauge},\,\lambda \rightarrow 0}
= S_0 + \lambda S_1 + O(\lambda^2) \,, 
\ee
where 
\be
S_{0} = - \int d^3 \sigma \frac{1}{2} \mathrm{tr}\,X =
- \int d^3 \sigma \frac{1}{2} \eta^{AB} \partial_A X^i \partial_B X^j \delta_{ij} \,,
\ee
describes 8 free bosons in three dimensions, with energy-momentum tensor
\be
T_{AB} = X_{AB} - \frac{1}{2} \eta_{AB} \mathrm{tr}\,X \,.
\ee
This energy-momentum tensor appears in the term linear in $\lambda$ in the expansion:
\be
\begin{split}
S_1 \equiv \frac{\partial S}{\partial \lambda}\Big|_{\lambda=0} &= \int d^3 \sigma \left( \frac{1}{4} \mathrm{tr}\, ( \eta^{-1} X \eta^{-1}X ) - \frac{1}{8} ( \mathrm{tr}\,X)^2\right)\\
& = \int d^3\sigma \left( \frac{1}{4} \mathrm{tr}\, ( \eta^{-1} T \eta^{-1} T ) - \frac{1}{4} ( \mathrm{tr}\,T)^2\right)\\
& = \int d^3\sigma \frac{1}{4}(\eta^{AB} \eta^{CD} - \eta^{AC} \eta^{BD} ) T_{AC} T_{BD} \,.
\end{split} 
\ee
In terms of a general three-dimensional metric $h_{AB}$, this would be consistent with a flow equation of the form
\be
\frac{\partial S}{\partial \lambda} = \int d^3\sigma \sqrt{|h|} \frac{1}{4}(h^{AB} h^{CD} - h^{AC} h^{BD} ) T_{AC} T_{BD} \,.
\label{suggestedflow}
\ee
as suggested in \cite{Bonelli:2018kik}.
(In two-dimensions, $( h^{AB} h^{CD} - h^{AC} h^{BD} ) = \det h^{-1} \epsilon^{AD} \epsilon^{BC}$ and hence we can rewrite this in terms of the determinant of the energy-momentum tensor.)
However, the full action \eqref{SM2static} does not obey the equation \eqref{suggestedflow}, nor any other potential expression (such as the suggestion of \cite{Taylor:2018xcy}) that we can identify -- see appendix \ref{flow}.
If the procedure we are following does produce an interesting deformation encoded in the M2 Dirac-Nambu-Goto action, more work is needed to uncover its structure.
Such a pursuit is beyond the immediate scope of this paper, so we instead turn to an analysis of this non-relativistic limit from the Hamiltonian perspective, in order to verify that a U-duality analogue of a TsT transformation connects the non-relativistic and relativistic geometries.

\subsection{Limits of the Hamiltonian action}
\label{hamm2}

We first need to split up the membrane time and spatial coordinates. To this end, we write $\sigma^A=(\tau,\sigma^\alpha)$, with $\alpha=1,2$. Define $\epsilon^{\alpha \beta} \equiv \epsilon^{0\alpha \beta}$ and $\{ X^\mu, X^\nu \} \equiv \epsilon^{0\alpha\beta} \partial_\alpha X^\mu \partial_\beta X^\nu$. 
The action in Hamiltonian form is
\be
S_{\text{M2}} = \int d^3 \sigma \left( \dot{X}^\mu P_\mu - u^\alpha(\mathcal{H}_u)_\alpha - e \mathcal{H}_e \right)\,,
\ee
where the constraints are
\be
(\mathcal{H}_u)_\alpha  = P_\mu \partial_\alpha X^\mu \,,
\ee
\be
\mathcal{H}_e = \frac{1}{2\TM} \begin{pmatrix}
\frac{1}{\sqrt{2}}\TM
\{ X^\mu, X^\nu \}
& P_\mu 
\end{pmatrix} 
\begin{pmatrix}
\frac{1}{\sqrt{2}^2} \left( 
2 g_{\mu[\nu} g_{\rho] \sigma} + C_{\mu\nu \kappa}g^{\kappa\lambda}C_{\rho\sigma\lambda}\right) & - \frac{1}{\sqrt{2}}
 C_{\mu\nu\kappa} g^{\kappa \rho} \\
-\frac{1}{\sqrt{2}}
 g^{\mu \kappa}C_{\kappa \rho\sigma} & g^{\mu\rho}
\end{pmatrix}
\begin{pmatrix}
\frac{1}{\sqrt{2}}\TM
\{ X^\rho, X^\sigma \} 
\\ 
P_\rho
\end{pmatrix}\,.
\ee
The background \eqref{M2GO} corresponds to
\be
g_{ab} = \lambda^{-2/3} \eta_{ab}\,,\quad g_{ij} = \lambda^{1/3} \delta_{ij} \,,\quad
C_{abc} = -\lambda^{-1} \epsilon_{abc}\,,
\ee
where $\epsilon_{012} =-1$.
Unlike in the string case, it is necessary to also rescale 
the Lagrange multiplier $e = \lambda^{1/3} \tilde e$, otherwise there is no contribution from $\mathcal{H}_e$ in the limit $\lambda \rightarrow 0$. 
Then,
\be
\begin{split}
e \mathcal{H}_e 
& = \frac{1}{2\TM} \tilde e
\begin{pmatrix}
\frac{1}{\sqrt{2}}
 \TM \{ X^a, X^b \} & P_a 
\end{pmatrix} 
\begin{pmatrix}
0 &   \frac{1}{\sqrt{2}} \epsilon_{abe} \eta^{ec} \\
 \frac{1}{\sqrt{2}} \epsilon_{cde}\eta^{ea} & \lambda \eta^{ac}
\end{pmatrix}
\begin{pmatrix}
\frac{1}{\sqrt{2}}
 \TM\{ X^c, X^d \} \\ P_c 
\end{pmatrix}
\\ &\qquad
+ \frac{1}{2\TM} \tilde e
\begin{pmatrix}
\frac{1}{\sqrt{2}}\TM
\{ X^i, X^j \} & P_i
\end{pmatrix} 
\begin{pmatrix}
 \lambda\delta_{i[k} \delta_{l]j} &  0 \\
0 &   \delta^{ij}
\end{pmatrix}
\begin{pmatrix}
\frac{1}{\sqrt{2}}\TM
\{ X^k, X^l \}\\ P_k
\end{pmatrix}\,.
\end{split}
\label{He}
\ee
Here we see a sum of two terms, the first involving the longitudinal part of the background, and the second involving the transverse directions.

The longitudinal part can be rewritten in terms of structures associated to the U-duality group acting in the $d=3$ dimensions, which is $\Gthree$.
To exhibit this structure, we view $( \frac{1}{\sqrt{2}}\{ X^a, X^b \} , P_a )$ as transforming in the $(\mathbf{3}, \mathbf{2})$ representation of $\Gthree$, i.e. more precisely $\epsilon_{abc} \frac{1}{\sqrt{2}} \{ X^b, X^c \}$ and $P_a$ each transform in the fundamental representation of $\mathrm{SL}(3)$ and together form a doublet under $\mathrm{SL}(2)$ U-duality transformations. 
We can describe these collectively by introducing a 6-dimensional multi-index $M$ such that $\mathcal{Z}_M \equiv ( \frac{1}{\sqrt{2}}\{ X^a, X^b \} , P_a )$.
Then we can write the longitudinal contribution to the Hamiltonian constraint as $\frac{\tilde e}{2T}\mathcal{Z}_M \gM^{MN} \mathcal{Z}_N$ in terms of a generalised metric combining the metric and three-form as follows:
\be
\gM^{MN} = |g|^{-1/6}
\begin{pmatrix}
g_{a[c} g_{d] b} + \frac{1}{2} C_{ab e}g^{ef}C_{cdf}& - 
\frac{1}{\sqrt{2}} C_{abe} g^{e c} \\
-\frac{1}{\sqrt{2}}
 g^{a e}C_{e cd} & g^{ac}
\end{pmatrix}\,.
\label{exgm}
\ee
For completeness, we discuss how this generalised metric factorises into separate $\mathrm{SL}(3)$ and $\mathrm{SL}(2)$ matrices in appendix \ref{morenonrie}.

For the M2 scaling limit background the generalised metric \eqref{M2GO} works out as:
\be
\gM^{MN} = \begin{pmatrix}
0 &  \frac{1}{\sqrt{2}}\epsilon_{abe} \eta^{ec} \\
 \frac{1}{\sqrt{2}}\epsilon_{cde}\eta^{ea} & \lambda \eta^{ac}
\end{pmatrix}
\ee
and for $\lambda \rightarrow 0$ the bottom right block corresponding to the inverse spacetime metric degenerates, in which case we can not describe a conventional relativistic geometry anymore.
Again though, we can factorise out the $\lambda$ dependence, now using a \emph{trivector} transformation:
\be
\gM^{MN} = U^M{}_K (\lambda) U^N{}_L (\lambda) \gM^{KL} (\lambda=0) \,,
\ee
where
\be
U^M{}_N(\lambda) = \begin{pmatrix}
\delta_{ab}^{[cd]} & 0 \\
\frac{1}{\sqrt{2}} \Omega^{acd}(\lambda) & \delta^a_c
\end{pmatrix}\,,\quad
\Omega^{abc}(\lambda) = \frac{\lambda}{2} \epsilon^{abc} \,.
\ee
is an element of $\Gthree$,
and
\be
\gM^{MN}(\lambda=0) = \begin{pmatrix}
0 & \frac{1}{\sqrt{2}} \epsilon_{abe} \eta^{ec} \\
 \frac{1}{\sqrt{2}}\epsilon_{cde}\eta^{ea} & 0
\end{pmatrix}
\label{GOMlimit}
\ee
is a non-relativistic parametrisation of an $\Gthree$ generalised metric.
So far this is entirely similar to the string theory limit.
As we would expect, the naive singularity of the background \eqref{M2GO} in the limit manifests itself as a degeneration of a block of the generalised metric appearing naturally in the Hamiltonian, and in place of  a bivector transformation (which are non-geometric counterparts of shifts of the string theory two-form) we have instead the appearance of a trivector (the non-geometric counterpart of the M-theory three-form) \cite{Malek:2013sp}.

What is less similar is the need to transform the worldvolume Lagrange multiplier $e$, although this is not so surprising as it is expected on general grounds to not be inert under duality transformations, reflecting a scaling of the worldvolume metric in the Polyakov formulation \cite{Duff:1990hn}. 

Furthermore, in the second line of equation \eqref{He} we see that a term involving solely the transverse coordinates $X^i$ vanishes for $\lambda \rightarrow 0$. How we could generate this term for $\lambda \neq 0$ using a $\Gthree$ transformation is a bit mysterious. This may involve the reformulation of the full membrane Hamiltonian in a ``duality covariant'' form in which the worldvolume tension is encoded in a charge vector in a representation of the U-duality group (ideas that have been explored in \cite{Sakatani:2017vbd,Arvanitakis:2018hfn} for various branes). One needs to contract two copies of this charge with the generalised metric to obtain a scalar tension, which may be what is vanishing when $\lambda =0$ and non-vanishing when $\lambda \neq 0$ \emph{if one keeps the charge vector fixed}. This raises subtle questions about the independence of the M2 and M5 actions under U-duality transformations (for $\Gthree$ the charge vector is a doublet corresponding in toroidal backgrounds to an unwrapped M2 and the M5 wrapped on three directions, and these transform into each other).
With an apology for ending on an unresolved issue which is interesting (to the author) but somewhat technical and far removed from the primary goals of this paper, we will now conclude our discussion with the intention of returning to this particular problem in a future work.

\subsection{Lessons and morals}

We see from this section that though some form of the structures we are investigating persist beyond string theory and into the eleven-dimensional realm of M-theory, there are as expected added difficulties when going from strings to branes in general, and the overall picture is much less clear.
Nonetheless we think it is interesting to make these comparisons.

As a final comment, we must emphasise that our approach throughout this paper has implicitly been largely driven by knowledge gained from the development of double and exceptional field theory, which formulate the $O(D,D)$ T- and $E_{d(d)}$ U-duality symmetries in a unified approach and in particular treat the generalised metric as an independent field (see \cite{Berman:2019biz} for a short conceptual introduction). 
It will be interesting to see if this unlikely confluence of topics -- non-relativistic string theory, $T \bar T$ deformations, and duality covariant formalisms -- can produce real insights into the space of physical theories.

\section*{Acknowledgements} 

I am supported by an FWO-Vlaanderen Postdoctoral Fellowship, and also in part by the FWO-Vlaanderen through the project G006119N and by the Vrije Universiteit Brussel through the Strategic Research Program ``High-Energy Physics''. I thank Alex Arvanitakis, Gerben Oling, Jeong-Hyuck Park, Alexander Sevrin and Daniel Thompson for helpful discussions.

\appendix

\section{Dirac-Nambu-Goto flow equation}
\label{flow}

Suppose we want to take a non-relativistic limit on a general $p$-brane (with a $p+1$ dimensional worldvolume).
Let's write flat spacetime in the form $ds^2 = \lambda^{-2/(p+1)} ( \eta_{ab} dX^a dX^b + \lambda \delta_{ij} dX^i dX^j)$, and take the $(p+1)$-form to which the brane couples to be given by $C_{a_1 \dots a_{p+1}} = - \lambda^{-1} \epsilon_{a_1 \dots a_{p+1}}$.
Let $A,B=0,\dots p$ be worldvolume indices, then the pullbacks of the metric and three-form are:
\be
g_{AB} = \lambda^{-2/(p+1)}  \left( h_{AB} + \lambda X_{AB} \right) \,,\quad
C_{A_1 \dots A_{p+1}} = - \lambda^{-1} \partial_{A_1} X^{a_1} \dots \partial_{A_{p+1}} X^{a_{p+1}} \epsilon_{a_1 \dots a_{p+1}} \,,
\ee
where $h_{AB} \equiv \partial_A X^a \partial_B X^b \eta_{ab}$, $X_{AB} \equiv \partial_A X^i \partial_B X^j \delta_{ij}$.

In principle, in string theory we may also need to consider a non-trivial dilaton as well.
Here, we are mainly interested in comparing the $p=1$ case of the usual string (in which case the above metric is string frame) and the $p=2$ case of the usual membrane in 11-dimensions (in which case the above metric is Einstein frame).

We play a trick with the $(p+1)$-form.
We write the determinant of $h_{AB}$ in the following manner:
\be
\begin{split}
\det h_{AB} & = \frac{1}{(p+1)!} \epsilon^{A_1 \dots A_{p+1}} \epsilon^{B_1 \dots B_{p+1}}
 h_{A_1 B_1} \dots h_{A_{p+1} B_{p+1}} 
 \\
 & = \frac{1}{(p+1)!} \epsilon^{A_1 \dots A_{p+1}} \partial_{A_1} X^{a_1} \dots \partial_{A_{p+1}} X^{a_{p+1}} \epsilon^{B_1 \dots B_{p+1}} \partial_{B_1} X^{b_1} \dots \partial_{B_{p+1}} X^{b_{p+1}}
 \eta_{a_1b_1} \dots \eta_{a_{p+1} b_{p+1}} \\
 & = - \left( \frac{1}{(p+1)!} \epsilon^{A_1 \dots A_{p+1}} \partial_{A_1} X^{a_1} \dots \partial_{A_{p+1}} X^{a_{p+1}} \epsilon_{a_1 \dots a_{p+1}} \right)^2\,.
\end{split}
\ee
We then see that the coupling to the $(p+1)$-form is $\frac{1}{(p+1)!} \epsilon^{A_1 \dots A_{p+1}} C_{A_1 \dots A_{p+1}} = -\frac{1}{\lambda} \sqrt{-\det h}$.
The Nambu-Goto action in this background can then be succinctly written as:
\be
S = \int d^{p+1} \sigma \frac{1}{\lambda} \left( \sqrt{-\det h} - \sqrt{-\det (h+\lambda X)} \right)\,.
\ee
(This recovers a form of the action used in \cite{Townsend:2019koy}.)
We see immediately that the $\lambda\rightarrow 0$ limit of this is
\be
S\Big|_{\lambda \rightarrow 0} = - \int d^{p+1}\sigma\sqrt{-\det h} \frac{1}{2} h^{AB} \partial_A X^i \partial_B X^j \delta_{ij} + O(\lambda)\,.
\ee
Let's now compute the dependence on $\lambda$.
We have: 	
\be
\begin{split}
\frac{\partial S}{\partial \lambda} &
=
\int d^{p+1} \sigma \frac{1}{\lambda^2}\left(  -\sqrt{-\det h} + \sqrt{ - \det ( h+ \lambda X) }\left( 1 -  
 \frac{1}{2}[(h+\lambda X)^{-1}]^{AB} \lambda X_{AB})\right)
\right)
\\ & 
=
\int d^{p+1}\sigma  \frac{\sqrt{-\det h} }{\lambda^2}\left(  -1 + \frac{\sqrt{ - \det ( h+ \lambda X) }}{\sqrt{-\det h} }\left( \frac{1-p}{2} 
+\frac{1}{2}[(h+\lambda X)^{-1}]^{AB}h_{AB})\right) 
\right) \,.
\end{split}
\ee
The appearance of the factor $1-p$ which vanishes for the string is possibly already worth noticing.

In order to obtain the energy-momentum tensor we vary the action with respect to $h_{AB}$:
\be
\delta S = \int d^3 \sigma \frac{\delta h^{AB} }{2\lambda}
\left(
 - \sqrt{-\det h} h_{AB} 
+ \sqrt{-\det(h+\lambda X)} [(h+\lambda X)^{-1}]^{CD} h_{CA} h_{DB}
\right) \,,
\ee
and define $T_{AB} = - \frac{2}{\sqrt{-\det h}} \frac{\delta S}{\delta h^{AB}}$, hence:
\be
T_{AB} = \frac{1}{\lambda} \left( h_{AB} -\frac{\sqrt{-\det(h+\lambda X)} }{\sqrt{-\det h}} [(h+\lambda X)^{-1}]^{CD} h_{CA} h_{DB}\right)\,.
\ee
We can compute
\be
h^{AB} T_{AB} =\frac{1}{\lambda} \left( p+1 -\frac{\sqrt{-\det(h+\lambda X)} }{\sqrt{-\det h}} [(h+\lambda X)^{-1}]^{AB} h_{AB} \right) \,,
\ee
\be
\begin{split}
(h^{AB} T_{AB})^2 
= \frac{1}{\lambda^2} \Big(&
(p+1)^2 - 2(p+1) \frac{\sqrt{-\det(h+\lambda X)} }{\sqrt{-\det h}} [(h+\lambda X)^{-1}]^{AB} h_{AB}
\\ &
+ \frac{\det (h+\lambda X)}{\det h} [(h+\lambda X)^{-1}]^{AB} [(h+\lambda X)^{-1}]^{CD} h_{AB}h_{CD}
\Big)\,,
\end{split}
\ee
and
\be
\begin{split}
h^{AB} h^{CD} T_{AC} T_{BD} = \frac{1}{\lambda^2} \Big( &
p+1  - 2 \frac{\sqrt{-\det(h+\lambda X)} }{\sqrt{-\det h}} [(h+\lambda X)^{-1}]^{AB} h_{AB}
\\ & 
+ \frac{\det (h+\lambda X)}{\det h} [(h+\lambda X)^{-1}]^{AB} [(h+\lambda X)^{-1}]^{CD} h_{AC}h_{BD}
\Big)\,,
\end{split}
\ee
such that
\be
\begin{split} 
\frac{1}{4}  & 
\sqrt{-\det h} (h^{AB} h^{CD} - h^{AC} h^{BD} ) T_{AC} T_{BD} 
\\ &
= \frac{\sqrt{-\det h}}{4\lambda^2} \Big(
- p(p+1)  + 2p \frac{\sqrt{-\det(h+\lambda X)} }{\sqrt{-\det h}} [(h+\lambda X)^{-1}]^{AB} h_{AB}
\\ & \qquad\qquad\qquad\qquad
-\frac{\det (h+\lambda X)}{\det h} [(h+\lambda X)^{-1}]^{AB} [(h+\lambda X)^{-1}]^{CD}( h_{AB}h_{CD} - h_{AC} h_{BD} )
\Big)\,.
\end{split}
\label{putativeTT}
\ee

\subsubsection*{Flow equation for strings}

When $p=1$, we have the identity
\be
h_{AB}h_{CD} - h_{AC} h_{BD} = \det h \epsilon_{AD} \epsilon_{BC}\,,
\label{matrixinv}
\ee
implying
\be
\begin{split}
[(h+\lambda X)^{-1}]^{AB}
[(h+\lambda X)^{-1}]^{CD}&( h_{AB}h_{CD} - h_{AC} h_{BD})
\\& 
 = \det h  [(h+\lambda X)^{-1}]^{AB}
[(h+\lambda X)^{-1}]^{CD} \epsilon_{AD} \epsilon_{BC}
\\ &
= 2 \frac{\det h}{\det (h+\lambda X)}\,,
\end{split}
\ee
and hence
\be
\begin{split} 
\frac{1}{4} 
\sqrt{-\det h} &(h^{AB} h^{CD} - h^{AC} h^{BD} )   T_{AC} T_{BD} 
\\ &
= \frac{\sqrt{-\det h}}{\lambda^2} \Big(
-1 + \frac{1}{2} \frac{\sqrt{-\det(h+\lambda X)} }{\sqrt{-\det h}} [(h+\lambda X)^{-1}]^{AB} h_{AB}
\Big)\,.
\end{split}
\ee
Then indeed we have
\be
\frac{\partial S}{\partial \lambda} = \int d^2\sigma \frac{1}{4}  
\sqrt{-\det h} (h^{AB} h^{CD} - h^{AC} h^{BD} ) T_{AC} T_{BD} 
\ee
which corresponds to the determinant of the energy-momentum tensor by using \eqref{matrixinv}.

\subsubsection*{Flow equation for membranes}

When $p=2$, matters are not quite so simple. In place of \eqref{matrixinv} we can use the matrix identity:
\be
2 h_{A[B} h_{C]D} = \det h \, \epsilon_{AD E} \epsilon_{BCF} h^{EF} \,.
\label{matrix3}
\ee
From this, for instance, we can write
\be
\begin{split}
((h+\lambda X)^{-1})^{AB} h_{AB}&
= \frac{1}{2} \frac{1}{\det ( h+\lambda X)} \epsilon^{ACD} \epsilon^{BEF} h_{AB} ( h+\lambda X)_{CE} (h+\lambda X)_{DF}
\\
 & 
 =
 \frac{1}{2}\frac{\det h}{\det (h+\lambda X)} (h^{CE} h^{FD} - h^{CF} h^{ED} ) ( h+\lambda X)_{CE} (h+\lambda X)_{DF}
 \\ &
=  \frac{1}{\det (I+\lambda h^{-1} X)} \left(
3 + 2 \lambda \mathrm{tr}\,h^{-1} X 
+ \frac{1}{2} \lambda^2\left( ( \mathrm{tr}\, h^{-1} X)^2 - \mathrm{tr} \,(h^{-1} X)^2 \right)
\right)  \,.
\end{split} 
\ee
For three-by-three matrices, we have explicitly that (here $I$ is the identity matrix):
\be
\det (I + \lambda h^{-1} X ) 
= 1 + \lambda \mathrm{tr} \,h^{-1} X + \frac{1}{2} \lambda^2 ( ( \mathrm{tr}\, h^{-1} X)^2 - \mathrm{tr} \,(h^{-1} X)^2 )+ \lambda^3 \det h^{-1} X \,,
\ee
We can then write down the explicit expression for the derivative of the action:
\be
\begin{split}
\frac{\partial S}{\partial \lambda} &
= \int d^3\sigma
\frac{\sqrt{-\det h} }{\lambda^2}\left(  -1 + \sqrt{ \det ( I + \lambda h^{-1} X )} \left( -\frac{1}{2} 
+\frac{1}{2}[(h+\lambda X)^{-1}]^{AB}h_{AB})\right)
\right)
\\ 
& =
\int d^3\sigma 
\frac{\sqrt{-\det h} }{\lambda^2}\left(  -1 -\frac{1}{2}
\sqrtbig{ 1 + \lambda \mathrm{tr} \,h^{-1} X + \frac{1}{2} \lambda^2 ( ( \mathrm{tr}\, h^{-1} X)^2 - \mathrm{tr} \,(h^{-1} X)^2 )+ \lambda^3 \det h^{-1} X }\right.
\\ & \qquad\qquad\qquad\qquad\qquad\qquad
\left.
+ \frac{1}{2} \frac{3 + 2 \lambda \mathrm{tr}\,h^{-1} X 
+ \frac{1}{2} \lambda^2\left( ( \mathrm{tr}\, h^{-1} X)^2 - \mathrm{tr} \,(h^{-1} X)^2 \right)
 }{\sqrtbig{ 1 + \lambda \mathrm{tr} \,h^{-1} X + \frac{1}{2} \lambda^2 ( ( \mathrm{tr}\, h^{-1} X)^2 - \mathrm{tr} \,(h^{-1} X)^2 )+ \lambda^3 \det h^{-1} X }}
\right)\,,
\end{split} 
\label{flowdS}
\ee
Meanwhile, we also have, from \eqref{putativeTT} and using \eqref{matrix3}
\be
\begin{split} 
\int d^3\sigma
\frac{1}{4}  & 
\sqrt{-\det h} (h^{AB} h^{CD} - h^{AC} h^{BD} ) T_{AC} T_{BD} 
\\ &
= \int d^3\sigma \frac{\sqrt{-\det h}}{4\lambda^2} \Big(
-6  + 4 \frac{\sqrt{-\det(h+\lambda X)} }{\sqrt{-\det h}} [(h+\lambda X)^{-1}]^{AB} h_{AB}
 - 2 h^{AB} ( h_{AB} + \lambda X_{AB} )
\Big)
\\ &
= \int d^3\sigma
\frac{\sqrt{-\det h}}{\lambda^2} \Big(
-3 - \frac{1}{2} \lambda \mathrm{tr}\,(h^{-1}X)   +  \sqrt{ \det ( I + \lambda h^{-1} X )} [(h+\lambda X)^{-1}]^{AB} h_{AB}
\Big)\,,
\end{split}
\ee
hence this term is equal to
\be
\int d^3\sigma 
\frac{\sqrt{-\det h}}{\lambda^2} \left(
-3 - \frac{1}{2} \lambda \mathrm{tr}\,(h^{-1}X) + 
 \frac{3 + 2 \lambda \mathrm{tr}\,h^{-1} X 
+ \frac{1}{2} \lambda^2\left( ( \mathrm{tr}\, h^{-1} X)^2 - \mathrm{tr} \,(h^{-1} X)^2 \right)
 }{\sqrtbig{ 1 + \lambda \mathrm{tr} \,h^{-1} X + \frac{1}{2} \lambda^2 ( ( \mathrm{tr}\, h^{-1} X)^2 - \mathrm{tr} \,(h^{-1} X)^2 )+ \lambda^3 \det h^{-1} X }}
\right) \,.
\label{flowTT}
\ee
The result \eqref{flowdS} is not equal to \eqref{flowTT}.
Indeed expanding for $\lambda$ small:
\be
\begin{split}
\frac{\partial S}{\partial \lambda}  & - \int d^3\sigma
\frac{1}{4}  
\sqrt{-\det h} (h^{AB} h^{CD} - h^{AC} h^{BD} ) T_{AC} T_{BD} 
\\ & 
= \frac{\lambda}{16} \left(
( \mathrm{tr} \,h^{-1}X)^3 - 2 ( \mathrm{tr} \,h^{-1}X) ( ( \mathrm{tr}\, h^{-1} X)^2 - \mathrm{tr} \,(h^{-1} X)^2 )
+ 8 \det (h^{-1} X)
\right) 
+ O(\lambda^2)\,.
\end{split}
\ee
Observe that they do in fact agree to zeroth order in $\lambda$, which recovers the result we found in section \ref{DNG} where we expanded that far and no further.

An alternative generalisation suggested in \cite{Taylor:2018xcy} would be to instead try $(h^{AB}h^{CD} - \frac{1}{p} h^{AC}h^{BD}) T_{AC} T_{BD}$ in $p+1$ dimensions. However, changing the coefficient of the second term immediately restores a non-zero term at zeroth order in $\lambda$, with $h^{AB}T_{AB} = - \frac{1}{2} \mathrm{tr}\,h^{-1}X + O(\lambda)$. We sadly conclude that we have not shed any useful light on the question of higher dimensional generalisations of $T \bar T$. 

\section{Non-Riemannian generalised metrics for $\Gthree$}
\label{morenonrie}

Here we provide some additional material with which to interpret the discussion in section \ref{hamm2}.
First, we should exhibit the factorisation of the $\Gthree$ generalised metric \eqref{exgm} into $\mathrm{SL}(3)$ and $\mathrm{SL}(2)$ factors.
Write $C_{abc} = C \epsilon_{abc}$ and note that $\epsilon_{abe} \epsilon_{cdf}g^{ef} = 2 \det g^{-1} \,g_{a[c}g_{d]b}$, $\epsilon^{abe} \epsilon^{cdf} g_{ac} g_{bd} = 2 \det g^{-1} g^{ef}$.
Then
\be
\begin{split}
|g|^{1/6}
\mathcal{Z}_M \gM^{MN} \mathcal{Z}_N & 
=
\begin{pmatrix}
\frac{1}{\sqrt{2}}
\{ X^a, X^b \} & P_a 
\end{pmatrix} 
\begin{pmatrix}
g_{a[c} g_{d] b} + \frac{1}{2} C_{ab e}g^{ef}C_{cdf}& - 
\frac{1}{\sqrt{2}} C_{abe} g^{e c} \\
-\frac{1}{\sqrt{2}}
 g^{a e}C_{e cd} & g^{ac}
\end{pmatrix}
\begin{pmatrix}
\frac{1}{\sqrt{2}}
\{ X^c, X^d \} \\ P_c 
\end{pmatrix}
\\ & =
\begin{pmatrix}
\frac{1}{2} \epsilon_{acd}
\{ X^c, X^d \} & P_a 
\end{pmatrix} 
\begin{pmatrix}
g^{ab} ( -| g|  + C^2) & 
-C g^{ab} \\
- C g^{ab} & g^{ab}
\end{pmatrix}
\begin{pmatrix}
\frac{1}{2} \epsilon_{bcd}
\{ X^c, X^d \} \\ P_c 
\end{pmatrix}
\end{split} 
\ee
and there is a factorisation $\gM^{MN} = \gM^{ab} \mathcal{H}^{\underline{\alpha} \underline{\beta}}$ into a three-by-three matrix transforming under $\mathrm{SL}(3)$ (corresponding to geometric coordinate transformations):
\be
\gM^{ab} \equiv |g|^{1/3} g^{ab}
\ee
and a two-by-two matrix transforming under $\mathrm{SL}(2)$ (corresponding to non-trivial U-duality transformations):
\be
 \mathcal{H}^{\underline{\alpha} \underline{\beta}} \equiv 
|g|^{-1/2} \begin{pmatrix} 
-|g| + C^2 & - C \\
- C & 1
\end{pmatrix}\,.
\ee
We introduced an $\mathrm{SL}(2)$ fundamental index $\underline{\alpha} = 1,2$ with the understanding that $\mathcal{Z}_{M} \equiv \mathcal{Z}_{a \underline{\alpha}}$ has components $\mathcal{Z}_{a1} =\frac{1}{\sqrt{2}} \epsilon_{abc} \{ X^b, X^c \}$, $\mathcal{Z}_{a2} = P_a$.

Perhaps somewhat unusually, $g_{ab}$ has Lorentzian signature here; our generalised metric parametrises a Lorentzian signature coset 
\cite{Hull:1998br}.

The paper \cite{Berman:2019izh} investigated some examples of non-Riemannian parametriations of the generalised metrics that are valued in, and transform under, the U-duality groups $E_{d(d)}$. A general classification such as is available for the $O(D,D)$ case was not provided.
We would therefore like to make the structure of the $\Gthree$ case more transparent here. 
Assuming that $\gM^{MN}$ and hence both $\gM^{ab}$ and $\mathcal{H}^{\underline\alpha\underline\beta}$ are invertible (both these blocks are needed to formulate the supergravity dynamics \cite{Hohm:2015xna}), and symmetric, then a general parametrisation of these factors will \emph{not} correspond to a Riemannian metric and three-form if 
\be
 \mathcal{H}^{\underline{\alpha} \underline{\beta}} 
= 
\begin{pmatrix}
a & \pm 1 \\ 
\pm 1 & 0
\end{pmatrix}  
\ee
which we remark necessarily has determinant $-1$. In any other case, we will be able to extract a definition of $|g| \neq 0$ from the bottom right entry, and then use the fact that $\gM^{ab}$ is necessarily a non-degenerate matrix to define $g^{ab}$.

The non-relativistic limit described by \eqref{GOMlimit} corresponds to
\be
\gM^{ab}= \eta^{ab} \,,\quad 
\mathcal{H}^{\underline{\alpha} \underline{\beta}} = \begin{pmatrix}
0 & 1 \\ 1 & 0 
\end{pmatrix}\,,
\ee
and the trivector deformation to the $\mathrm{SL}(2)$ transformation
\be
U^{\underline{\alpha}}{}_{\underline{\beta}} = \begin{pmatrix} 1 & 0 \\ \frac{\lambda}{2} & 1 \end{pmatrix} \,.
\ee

\section{Buscher dualities of the generalised metric}
\label{buscher}

A convenient way to describe a single radius inversion T-duality, or Buscher transformation \cite{Buscher:1987sk, Buscher:1987qj}, along one direction in $O(D,D)$ language is to introduce a $D$-dimensional (constant) vector $n^\mu$ and dual covector $n_\mu$ such that $n^\mu n_\mu =1$.
Then dualising in the direction $n^\mu$ corresponds to the following $O(D,D)$ transformation:
\be
(\mathcal{T}_n)_M{}^N = 
\begin{pmatrix}
\delta_\mu^\nu - n_\mu n^\nu & n_\mu n_\nu \\
n^\mu n^\nu & \delta^\mu_\nu - n^\mu n_\nu 
\end{pmatrix} \,,
\ee
which is its own inverse.
If we choose coordinates such that $X^\mu =( X^i, X^z)$ with $X^z$ the isometry direction, then we can take $n^\mu = \delta^\mu_z$, $n_\mu = \delta_\mu^z$. 
In this case, acting on the generalised metric we have
\be
\mathcal{H}_{MN} \rightarrow \widetilde{\mathcal{H}}_{MN} = ( \mathcal{T}_z)_M{}^K (\mathcal{T}_z)_N{}^L \mathcal{H}_{KL}\,,
\ee
and in component language this is just a permutation swapping the components with upper $z$ for the components with lower $z$, and vice versa, hence:
\be
\widetilde{\mathcal{H}}_{ij} = \mathcal{H}_{ij}\,,\quad
\widetilde{\mathcal{H}}_{iz} = \mathcal{H}_{i}{}^z\,,\quad
\widetilde{\mathcal{H}}_{zz} = \mathcal{H}^{zz}\,,
\ee
\be
\widetilde{\mathcal{H}}_{i}{}^{j} = \mathcal{H}_{i}{}^{j}\,,\quad
\widetilde{\mathcal{H}}_{i}{}^{z} = \mathcal{H}_{iz}\,,\quad
\widetilde{\mathcal{H}}_{z}{}^{z} = \mathcal{H}_{z}{}^z\,,\quad
\widetilde{\mathcal{H}}_{z}{}^{i} = \mathcal{H}^{zi}\,,\quad
\ee
\be
\widetilde{\mathcal{H}}^{ij} = \mathcal{H}^{ij}\,,\quad
\widetilde{\mathcal{H}}^{iz} = \mathcal{H}^{i}{}_z\,,\quad
\widetilde{\mathcal{H}}^{zz} = \mathcal{H}_{zz}\,.
\ee
Parametrised in terms of $g$ and $B$ this reproduces the usual Buscher rules; starting with $\tilde g_{zz} = \frac{1}{g_{zz}}$.
However, even if $g_{zz} = 0$, i.e. we have a null isometry, we can consider safely the transformation of the generalised metric if we interpret the resulting $\widetilde{\mathcal{H}}_{MN}$ as parametrising a non-relativistic geometry. 

\section{Effective tension of the non-relativistic limit}
\label{unitz}

In \cite{Gomis:2000bd}, the non-relativistic limit is taken starting from the Polyakov action
\be
S = - \frac{1}{4\pi \alpha^\prime} \int d^2\sigma \left( 
\eta_{ab} \partial_\alpha X^a \partial^\alpha X^b + \frac{\alpha^\prime}{\alpha^\prime_{\text{eff}}} \delta_{ij} \partial_\alpha X^i \partial^\alpha X^j 
+ 2 \left(1 - \frac{\alpha^\prime}{2\alpha^\prime_{\text{eff}}}\right) \epsilon^{\alpha \beta} \partial_\alpha X^0 \partial_\alpha X^1
\right) \,,
\ee
and sending $\alpha^\prime \rightarrow 0$.
If we pull out an overall factor of $\alpha^\prime/\alpha^\prime_{\text{eff}}$, this action is:
\be
S = - \frac{1}{4\pi \alpha^\prime_{\text{eff}} }\int d^2\sigma \left( 
\frac{\alpha^\prime_{\text{eff}}}{\alpha^\prime} \eta_{ab} \partial_\alpha X^a \partial^\alpha X^b +  \delta_{ij} \partial_\alpha X^i \partial^\alpha X^j 
+ 2 \left( \frac{\alpha^\prime_{\text{eff}}}{\alpha^\prime} - \frac{1}{2}\right) \epsilon^{\alpha \beta} \partial_\alpha X^0 \partial_\alpha X^1
\right) \,,
\label{SGOresc}
\ee
We can now identify
\be
\omega^2 \equiv \frac{\alpha^\prime_{\text{eff}}}{\alpha^\prime} \,,
\ee
and view this in the form of our initial background \eqref{initial}. The $B$-field is $B_{01} = \omega^2 - \mu$, with $\mu=1/2$ in \cite{Gomis:2000bd}.

The dilaton or string coupling of \cite{Gomis:2000bd} was taken to be $e^\phi = e^{\phi_{\text{eff}}} \sqrt{ \frac{\alpha^\prime_{\text{eff}}}{\alpha^\prime}}$.
Hence, using the metric in \eqref{SGOresc}, the T-duality invariant dilaton $d$ defined by $e^{-2d} = e^{-2\phi} \sqrt{-\det g} = e^{-2\phi_{\text{eff}}}$ is invariant in the scaling limit.

In the main text of this paper, we set the effective tension, $T_{\text{eff}}= \frac{1}{2\pi \alpha^\prime_{\text{eff}}}$ to one.
Had we not, the Nambu-Goto action should have been
\be
S_{\text{NG}} = - T_{\text{eff}} \int d^2 \sigma L_{\text{NG}} \,,
\ee
where $L_{\text{NG}}$ is the Nambu-Goto Lagrangian (with $B$-field) appearing in \eqref{NG}.
Then defining the energy-momentum tensor simply by the usual metric variation, it would naturally come with a factor of $T_{\text{eff}}$, and hence the determinant with a factor of $T_{\text{eff}}^2$, whereas the derivative of $S_{\text{NG}}$ with respect to $\lambda$ would still only carry a single factor of $T_{\text{eff}}$.
The flow equation would then be:
\be
\frac{\partial S_{\text{NG}}}{\partial \lambda} = 
\frac{1}{T_{\text{eff}}} \int d^2\sigma \frac{1}{2} \det (T_{\alpha\beta})\,.
\ee
If we define a dimensionful $T \bar T$ parameter, with units of length squared, by $\tilde \lambda = \lambda / T_{\text{eff}}$, we get
\be
\frac{\partial S_{\text{NG}}}{\partial \tilde\lambda} = 
\int d^2\sigma \frac{1}{2} \det (T_{\alpha\beta})\,,
\ee
and $\tilde\lambda = 2\pi \alpha^\prime$.

\bibliography{CurrentBib}

\providecommand{\href}[2]{#2}\begingroup\raggedright\begin{thebibliography}{10}

\bibitem{Zamolodchikov:2004ce}
A.~B. Zamolodchikov, \emph{{Expectation value of composite field T anti-T in
  two-dimensional quantum field theory}},
  \href{https://arxiv.org/abs/hep-th/0401146}{{\ttfamily hep-th/0401146}}.

\bibitem{Smirnov:2016lqw}
F.~A. Smirnov and A.~B. Zamolodchikov, \emph{{On space of integrable quantum
  field theories}},
  \href{https://doi.org/10.1016/j.nuclphysb.2016.12.014}{\emph{Nucl. Phys.}
  {\bfseries B915} (2017) 363--383},
  [\href{https://arxiv.org/abs/1608.05499}{{\ttfamily 1608.05499}}].

\bibitem{Cavaglia:2016oda}
A.~Cavaglià, S.~Negro, I.~M. Szécsényi and R.~Tateo, \emph{{$T
  \bar{T}$-deformed 2D Quantum Field Theories}},
  \href{https://doi.org/10.1007/JHEP10(2016)112}{\emph{JHEP} {\bfseries 10}
  (2016) 112}, [\href{https://arxiv.org/abs/1608.05534}{{\ttfamily
  1608.05534}}].

\bibitem{Jiang:2019hxb}
Y.~Jiang, \emph{{Lectures on solvable irrelevant deformations of 2d quantum
  field theory}},  \href{https://arxiv.org/abs/1904.13376}{{\ttfamily
  1904.13376}}.

\bibitem{Bonelli:2018kik}
G.~Bonelli, N.~Doroud and M.~Zhu, \emph{{$T \bar{T}$-deformations in closed
  form}}, \href{https://doi.org/10.1007/JHEP06(2018)149}{\emph{JHEP} {\bfseries
  06} (2018) 149}, [\href{https://arxiv.org/abs/1804.10967}{{\ttfamily
  1804.10967}}].

\bibitem{Gomis:2000bd}
J.~Gomis and H.~Ooguri, \emph{{Nonrelativistic closed string theory}},
  \href{https://doi.org/10.1063/1.1372697}{\emph{J. Math. Phys.} {\bfseries 42}
  (2001) 3127--3151}, [\href{https://arxiv.org/abs/hep-th/0009181}{{\ttfamily
  hep-th/0009181}}].

\bibitem{Danielsson:2000gi}
U.~H. Danielsson, A.~Guijosa and M.~Kruczenski, \emph{{IIA/B, wound and
  wrapped}}, \href{https://doi.org/10.1088/1126-6708/2000/10/020}{\emph{JHEP}
  {\bfseries 10} (2000) 020},
  [\href{https://arxiv.org/abs/hep-th/0009182}{{\ttfamily hep-th/0009182}}].

\bibitem{Danielsson:2000mu}
U.~H. Danielsson, A.~Guijosa and M.~Kruczenski, \emph{{Newtonian gravitons and
  d-brane collective coordinates in wound string theory}},
  \href{https://doi.org/10.1088/1126-6708/2001/03/041}{\emph{JHEP} {\bfseries
  03} (2001) 041}, [\href{https://arxiv.org/abs/hep-th/0012183}{{\ttfamily
  hep-th/0012183}}].

\bibitem{Andringa:2012uz}
R.~Andringa, E.~Bergshoeff, J.~Gomis and M.~de~Roo, \emph{{'Stringy'
  Newton-Cartan Gravity}},
  \href{https://doi.org/10.1088/0264-9381/29/23/235020}{\emph{Class. Quant.
  Grav.} {\bfseries 29} (2012) 235020},
  [\href{https://arxiv.org/abs/1206.5176}{{\ttfamily 1206.5176}}].

\bibitem{Gomis:2004pw}
J.~Gomis, K.~Kamimura and P.~K. Townsend, \emph{{Non-relativistic
  superbranes}},
  \href{https://doi.org/10.1088/1126-6708/2004/11/051}{\emph{JHEP} {\bfseries
  11} (2004) 051}, [\href{https://arxiv.org/abs/hep-th/0409219}{{\ttfamily
  hep-th/0409219}}].

\bibitem{Lunin:2005jy}
O.~Lunin and J.~M. Maldacena, \emph{{Deforming field theories with U(1) x U(1)
  global symmetry and their gravity duals}},
  \href{https://doi.org/10.1088/1126-6708/2005/05/033}{\emph{JHEP} {\bfseries
  05} (2005) 033}, [\href{https://arxiv.org/abs/hep-th/0502086}{{\ttfamily
  hep-th/0502086}}].

\bibitem{Baggio:2018gct}
M.~Baggio and A.~Sfondrini, \emph{{Strings on NS-NS Backgrounds as Integrable
  Deformations}}, \href{https://doi.org/10.1103/PhysRevD.98.021902}{\emph{Phys.
  Rev.} {\bfseries D98} (2018) 021902},
  [\href{https://arxiv.org/abs/1804.01998}{{\ttfamily 1804.01998}}].

\bibitem{Araujo:2018rho}
T.~Araujo, E.~Colgáin, Y.~Sakatani, M.~M. Sheikh-Jabbari and H.~Yavartanoo,
  \emph{{Holographic integration of $T \bar{T}$ \& $J \bar{T}$ via $O(d,d)$}},
  \href{https://doi.org/10.1007/JHEP03(2019)168}{\emph{JHEP} {\bfseries 03}
  (2019) 168}, [\href{https://arxiv.org/abs/1811.03050}{{\ttfamily
  1811.03050}}].

\bibitem{Frolov:2019nrr}
S.~Frolov, \emph{{TTbar deformation and the light-cone gauge}},
  \href{https://arxiv.org/abs/1905.07946}{{\ttfamily 1905.07946}}.

\bibitem{Frolov:2019xzi}
S.~Frolov, \emph{{$T{\overline T}$, $\widetilde JJ$, $JT$ and $\widetilde JT$
  deformations}}, \href{https://doi.org/10.1088/1751-8121/ab581b}{\emph{J.
  Phys.} {\bfseries A53} (2020) 025401},
  [\href{https://arxiv.org/abs/1907.12117}{{\ttfamily 1907.12117}}].

\bibitem{Sfondrini:2019smd}
A.~Sfondrini and S.~J. van Tongeren, \emph{{$T\bar{T}$ deformations as $TsT$
  transformations}},
  \href{https://doi.org/10.1103/PhysRevD.101.066022}{\emph{Phys. Rev. D}
  {\bfseries 101} (2020) 066022},
  [\href{https://arxiv.org/abs/1908.09299}{{\ttfamily 1908.09299}}].

\bibitem{Jorjadze:2020ili}
G.~Jorjadze and S.~Theisen, \emph{{Canonical maps and integrability in $T\bar
  T$ deformed 2d CFTs}},  \href{https://arxiv.org/abs/2001.03563}{{\ttfamily
  2001.03563}}.

\bibitem{Harmark:2017rpg}
T.~Harmark, J.~Hartong and N.~A. Obers, \emph{{Nonrelativistic strings and
  limits of the AdS/CFT correspondence}},
  \href{https://doi.org/10.1103/PhysRevD.96.086019}{\emph{Phys. Rev.}
  {\bfseries D96} (2017) 086019},
  [\href{https://arxiv.org/abs/1705.03535}{{\ttfamily 1705.03535}}].

\bibitem{Harmark:2018cdl}
T.~Harmark, J.~Hartong, L.~Menculini, N.~A. Obers and Z.~Yan, \emph{{Strings
  with Non-Relativistic Conformal Symmetry and Limits of the AdS/CFT
  Correspondence}}, \href{https://doi.org/10.1007/JHEP11(2018)190}{\emph{JHEP}
  {\bfseries 11} (2018) 190},
  [\href{https://arxiv.org/abs/1810.05560}{{\ttfamily 1810.05560}}].

\bibitem{Bergshoeff:2018yvt}
E.~Bergshoeff, J.~Gomis and Z.~Yan, \emph{{Nonrelativistic String Theory and
  T-Duality}}, \href{https://doi.org/10.1007/JHEP11(2018)133}{\emph{JHEP}
  {\bfseries 11} (2018) 133},
  [\href{https://arxiv.org/abs/1806.06071}{{\ttfamily 1806.06071}}].

\bibitem{Harmark:2019upf}
T.~Harmark, J.~Hartong, L.~Menculini, N.~A. Obers and G.~Oling, \emph{{Relating
  non-relativistic string theories}},
  \href{https://doi.org/10.1007/JHEP11(2019)071}{\emph{JHEP} {\bfseries 11}
  (2019) 071}, [\href{https://arxiv.org/abs/1907.01663}{{\ttfamily
  1907.01663}}].

\bibitem{Lee:2013hma}
K.~Lee and J.-H. Park, \emph{{Covariant action for a string in "doubled yet
  gauged" spacetime}},
  \href{https://doi.org/10.1016/j.nuclphysb.2014.01.003}{\emph{Nucl. Phys.}
  {\bfseries B880} (2014) 134--154},
  [\href{https://arxiv.org/abs/1307.8377}{{\ttfamily 1307.8377}}].

\bibitem{Ko:2015rha}
S.~M. Ko, C.~Melby-Thompson, R.~Meyer and J.-H. Park, \emph{{Dynamics of
  Perturbations in Double Field Theory \& Non-Relativistic String Theory}},
  \href{https://doi.org/10.1007/JHEP12(2015)144}{\emph{JHEP} {\bfseries 12}
  (2015) 144}, [\href{https://arxiv.org/abs/1508.01121}{{\ttfamily
  1508.01121}}].

\bibitem{Morand:2017fnv}
K.~Morand and J.-H. Park, \emph{{Classification of non-Riemannian
  doubled-yet-gauged spacetime}},
  \href{https://doi.org/10.1140/epjc/s10052-017-5257-z,
  10.1140/epjc/s10052-018-6394-8}{\emph{Eur. Phys. J.} {\bfseries C77} (2017)
  685}, [\href{https://arxiv.org/abs/1707.03713}{{\ttfamily 1707.03713}}].

\bibitem{Townsend:1999hi}
P.~K. Townsend, \emph{{Brane theory solitons}},  in \emph{{Proceedings, NATO
  Advanced Study Institute on Progress in String Theory and M-Theory: Cargese,
  France, May 24-June 5, 1999}}, pp.~265--296, 1999,
  \href{https://arxiv.org/abs/hep-th/0004039}{{\ttfamily hep-th/0004039}}.

\bibitem{Giveon:2017nie}
A.~Giveon, N.~Itzhaki and D.~Kutasov, \emph{{$ \mathrm{T}\overline{\mathrm{T}}
  $ and LST}}, \href{https://doi.org/10.1007/JHEP07(2017)122}{\emph{JHEP}
  {\bfseries 07} (2017) 122},
  [\href{https://arxiv.org/abs/1701.05576}{{\ttfamily 1701.05576}}].

\bibitem{Apolo:2018qpq}
L.~Apolo and W.~Song, \emph{{Strings on warped AdS$_{3}$ via $
  \mathrm{T}\bar{\mathrm{J}} $ deformations}},
  \href{https://doi.org/10.1007/JHEP10(2018)165}{\emph{JHEP} {\bfseries 10}
  (2018) 165}, [\href{https://arxiv.org/abs/1806.10127}{{\ttfamily
  1806.10127}}].

\bibitem{Apolo:2019zai}
L.~Apolo, S.~Detournay and W.~Song, \emph{{TsT, $T\bar{T}$ and black strings}},
   \href{https://arxiv.org/abs/1911.12359}{{\ttfamily 1911.12359}}.

\bibitem{Gomis:2019zyu}
J.~Gomis, J.~Oh and Z.~Yan, \emph{{Nonrelativistic String Theory in Background
  Fields}}, \href{https://doi.org/10.1007/JHEP10(2019)101}{\emph{JHEP}
  {\bfseries 10} (2019) 101},
  [\href{https://arxiv.org/abs/1905.07315}{{\ttfamily 1905.07315}}].

\bibitem{Gallegos:2019icg}
A.~D. Gallegos, U.~Gürsoy and N.~Zinnato, \emph{{Torsional Newton Cartan
  gravity from non-relativistic strings}},
  \href{https://arxiv.org/abs/1906.01607}{{\ttfamily 1906.01607}}.

\bibitem{Bergshoeff:2019pij}
E.~A. Bergshoeff, J.~Gomis, J.~Rosseel, C.~Şimşek and Z.~Yan, \emph{{String
  Theory and String Newton-Cartan Geometry}},
  \href{https://doi.org/10.1088/1751-8121/ab56e9}{\emph{J. Phys.} {\bfseries
  A53} (2020) 014001}, [\href{https://arxiv.org/abs/1907.10668}{{\ttfamily
  1907.10668}}].

\bibitem{Cho:2019ofr}
K.~Cho and J.-H. Park, \emph{{Remarks on the non-Riemannian sector in Double
  Field Theory}},
  \href{https://doi.org/10.1140/epjc/s10052-020-7648-9}{\emph{Eur.\ Phys.\ J.\
  C} {\bfseries 80} (2020) 101},
  [\href{https://arxiv.org/abs/1909.10711}{{\ttfamily 1909.10711}}].

\bibitem{Seiberg:2000ms}
N.~Seiberg, L.~Susskind and N.~Toumbas, \emph{{Strings in background electric
  field, space / time noncommutativity and a new noncritical string theory}},
  \href{https://doi.org/10.1088/1126-6708/2000/06/021}{\emph{JHEP} {\bfseries
  06} (2000) 021}, [\href{https://arxiv.org/abs/hep-th/0005040}{{\ttfamily
  hep-th/0005040}}].

\bibitem{Gopakumar:2000na}
R.~Gopakumar, J.~M. Maldacena, S.~Minwalla and A.~Strominger, \emph{{S duality
  and noncommutative gauge theory}},
  \href{https://doi.org/10.1088/1126-6708/2000/06/036}{\emph{JHEP} {\bfseries
  06} (2000) 036}, [\href{https://arxiv.org/abs/hep-th/0005048}{{\ttfamily
  hep-th/0005048}}].

\bibitem{Conti:2018jho}
R.~Conti, L.~Iannella, S.~Negro and R.~Tateo, \emph{{Generalised Born-Infeld
  models, Lax operators and the $ \mathrm{T}\overline{\mathrm{T}} $
  perturbation}}, \href{https://doi.org/10.1007/JHEP11(2018)007}{\emph{JHEP}
  {\bfseries 11} (2018) 007},
  [\href{https://arxiv.org/abs/1806.11515}{{\ttfamily 1806.11515}}].

\bibitem{Chang:2018dge}
C.-K. Chang, C.~Ferko and S.~Sethi, \emph{{Supersymmetry and $ T\overline{T} $
  deformations}}, \href{https://doi.org/10.1007/JHEP04(2019)131}{\emph{JHEP}
  {\bfseries 04} (2019) 131},
  [\href{https://arxiv.org/abs/1811.01895}{{\ttfamily 1811.01895}}].

\bibitem{Brennan:2019azg}
T.~D. Brennan, C.~Ferko and S.~Sethi, \emph{{A Non-Abelian Analogue of DBI from
  $T \overline{T}$}},  \href{https://arxiv.org/abs/1912.12389}{{\ttfamily
  1912.12389}}.

\bibitem{Blair:2016xnn}
C.~D.~A. Blair, \emph{{Doubled strings, negative strings and null waves}},
  \href{https://doi.org/10.1007/JHEP11(2016)042}{\emph{JHEP} {\bfseries 11}
  (2016) 042}, [\href{https://arxiv.org/abs/1608.06818}{{\ttfamily
  1608.06818}}].

\bibitem{Park:2015bza}
J.-H. Park, S.-J. Rey, W.~Rim and Y.~Sakatani, \emph{{O(D, D) covariant Noether
  currents and global charges in double field theory}},
  \href{https://doi.org/10.1007/JHEP11(2015)131}{\emph{JHEP} {\bfseries 11}
  (2015) 131}, [\href{https://arxiv.org/abs/1507.07545}{{\ttfamily
  1507.07545}}].

\bibitem{Dijkgraaf:2016lym}
R.~Dijkgraaf, B.~Heidenreich, P.~Jefferson and C.~Vafa, \emph{{Negative Branes,
  Supergroups and the Signature of Spacetime}},
  \href{https://doi.org/10.1007/JHEP02(2018)050}{\emph{JHEP} {\bfseries 02}
  (2018) 050}, [\href{https://arxiv.org/abs/1603.05665}{{\ttfamily
  1603.05665}}].

\bibitem{LevyLeblondCarroll}
J.-M. L\'evy-Leblond, \emph{Une nouvelle limite non-relativiste du groupe de
  poincar\'e}, {\emph{Annales de l'I.H.P. Physique th\'eorique} {\bfseries 3}
  (1965) 1--12}.

\bibitem{Duval:2014uoa}
C.~Duval, G.~W. Gibbons, P.~A. Horvathy and P.~M. Zhang, \emph{{Carroll versus
  Newton and Galilei: two dual non-Einsteinian concepts of time}},
  \href{https://doi.org/10.1088/0264-9381/31/8/085016}{\emph{Class. Quant.
  Grav.} {\bfseries 31} (2014) 085016},
  [\href{https://arxiv.org/abs/1402.0657}{{\ttfamily 1402.0657}}].

\bibitem{Townsend:2019koy}
P.~K. Townsend, \emph{{A manifestly Lorentz invariant chiral boson action}},
  \href{https://arxiv.org/abs/1912.04773}{{\ttfamily 1912.04773}}.

\bibitem{Chakrabarti:2020pxr}
S.~Chakrabarti and M.~Raman, \emph{{Chiral Decoupling from Irrelevant
  Deformations}},  \href{https://arxiv.org/abs/2001.06870}{{\ttfamily
  2001.06870}}.

\bibitem{Gomis:2005pg}
J.~Gomis, J.~Gomis and K.~Kamimura, \emph{{Non-relativistic superstrings: A New
  soluble sector of AdS(5) x S**5}},
  \href{https://doi.org/10.1088/1126-6708/2005/12/024}{\emph{JHEP} {\bfseries
  12} (2005) 024}, [\href{https://arxiv.org/abs/hep-th/0507036}{{\ttfamily
  hep-th/0507036}}].

\bibitem{Roychowdhury:2019vzh}
D.~Roychowdhury, \emph{{On integrability in nonrelativistic string theory}},
  \href{https://arxiv.org/abs/1904.06485}{{\ttfamily 1904.06485}}.

\bibitem{Park:2016sbw}
J.-H. Park, \emph{{Green-Schwarz superstring on doubled-yet-gauged spacetime}},
  \href{https://doi.org/10.1007/JHEP11(2016)005}{\emph{JHEP} {\bfseries 11}
  (2016) 005}, [\href{https://arxiv.org/abs/1609.04265}{{\ttfamily
  1609.04265}}].

\bibitem{Blair:2019qwi}
C.~D.~A. Blair, \emph{{A worldsheet supersymmetric Newton-Cartan string}},
  \href{https://doi.org/10.1007/JHEP10(2019)266}{\emph{JHEP} {\bfseries 10}
  (2019) 266}, [\href{https://arxiv.org/abs/1908.00074}{{\ttfamily
  1908.00074}}].

\bibitem{Baggio:2018rpv}
M.~Baggio, A.~Sfondrini, G.~Tartaglino-Mazzucchelli and H.~Walsh, \emph{{On $
  T\overline{T} $ deformations and supersymmetry}},
  \href{https://doi.org/10.1007/JHEP06(2019)063}{\emph{JHEP} {\bfseries 06}
  (2019) 063}, [\href{https://arxiv.org/abs/1811.00533}{{\ttfamily
  1811.00533}}].

\bibitem{Cardy:2018sdv}
J.~Cardy, \emph{{The $ T\overline{T} $ deformation of quantum field theory as
  random geometry}}, \href{https://doi.org/10.1007/JHEP10(2018)186}{\emph{JHEP}
  {\bfseries 10} (2018) 186},
  [\href{https://arxiv.org/abs/1801.06895}{{\ttfamily 1801.06895}}].

\bibitem{Taylor:2018xcy}
M.~Taylor, \emph{{TT deformations in general dimensions}},
  \href{https://arxiv.org/abs/1805.10287}{{\ttfamily 1805.10287}}.

\bibitem{Berman:2019izh}
D.~S. Berman, C.~D.~A. Blair and R.~Otsuki, \emph{{Non-Riemannian geometry of
  M-theory}}, \href{https://doi.org/10.1007/JHEP07(2019)175}{\emph{JHEP}
  {\bfseries 07} (2019) 175},
  [\href{https://arxiv.org/abs/1902.01867}{{\ttfamily 1902.01867}}].

\bibitem{Malek:2013sp}
E.~Malek, \emph{{Timelike U-dualities in Generalised Geometry}},
  \href{https://doi.org/10.1007/JHEP11(2013)185}{\emph{JHEP} {\bfseries 11}
  (2013) 185}, [\href{https://arxiv.org/abs/1301.0543}{{\ttfamily 1301.0543}}].

\bibitem{Duff:1990hn}
M.~J. Duff and J.~X. Lu, \emph{{Duality Rotations in Membrane Theory}},
  \href{https://doi.org/10.1016/0550-3213(90)90565-U}{\emph{Nucl. Phys.}
  {\bfseries B347} (1990) 394--419}.

\bibitem{Sakatani:2017vbd}
Y.~Sakatani and S.~Uehara, \emph{{Exceptional M-brane sigma models and
  $\eta$-symbols}}, \href{https://doi.org/10.1093/ptep/pty021}{\emph{PTEP}
  {\bfseries 2018} (2018) 033B05},
  [\href{https://arxiv.org/abs/1712.10316}{{\ttfamily 1712.10316}}].

\bibitem{Arvanitakis:2018hfn}
A.~S. Arvanitakis and C.~D.~A. Blair, \emph{{The Exceptional Sigma Model}},
  \href{https://doi.org/10.1007/JHEP04(2018)064}{\emph{JHEP} {\bfseries 04}
  (2018) 064}, [\href{https://arxiv.org/abs/1802.00442}{{\ttfamily
  1802.00442}}].

\bibitem{Berman:2019biz}
D.~S. Berman, \emph{{A Kaluza–Klein Approach to Double and Exceptional Field
  Theory}}, \href{https://doi.org/10.1002/prop.201910002}{\emph{Fortsch. Phys.}
  {\bfseries 67} (2019) 1910002},
  [\href{https://arxiv.org/abs/1903.02860}{{\ttfamily 1903.02860}}].

\bibitem{Hull:1998br}
C.~M. Hull and B.~Julia, \emph{{Duality and moduli spaces for timelike
  reductions}},
  \href{https://doi.org/10.1016/S0550-3213(98)00519-7}{\emph{Nucl. Phys.}
  {\bfseries B534} (1998) 250--260},
  [\href{https://arxiv.org/abs/hep-th/9803239}{{\ttfamily hep-th/9803239}}].

\bibitem{Hohm:2015xna}
O.~Hohm and Y.-N. Wang, \emph{{Tensor hierarchy and generalized Cartan calculus
  in SL(3) x SL(2) exceptional field theory}},
  \href{https://doi.org/10.1007/JHEP04(2015)050}{\emph{JHEP} {\bfseries 04}
  (2015) 050}, [\href{https://arxiv.org/abs/1501.01600}{{\ttfamily
  1501.01600}}].

\bibitem{Buscher:1987sk}
T.~H. Buscher, \emph{{A Symmetry of the String Background Field Equations}},
  \href{https://doi.org/10.1016/0370-2693(87)90769-6}{\emph{Phys. Lett.}
  {\bfseries B194} (1987) 59--62}.

\bibitem{Buscher:1987qj}
T.~H. Buscher, \emph{{Path Integral Derivation of Quantum Duality in Nonlinear
  Sigma Models}},
  \href{https://doi.org/10.1016/0370-2693(88)90602-8}{\emph{Phys. Lett.}
  {\bfseries B201} (1988) 466--472}.

\end{thebibliography}\endgroup

\end{document}